\documentclass[useAMS, usenatbib]{mn2e}
\usepackage{times}
\usepackage{graphicx}
\usepackage{subfigure}
\usepackage{psfrag}
\usepackage{amssymb, amsmath}

\def\reff@jnl#1{{\rm#1\/}}
\def\aj{\reff@jnl{AJ}}                 
\def\araa{\reff@jnl{ARA\&A}}           
\def\apj{\reff@jnl{ApJ}}               
\def\apjl{\reff@jnl{ApJ}}              
\def\apjs{\reff@jnl{ApJS}}             
\def\ao{\reff@jnl{Appl.Optics}}        
\def\apss{\reff@jnl{Ap\&SS}}           
\def\aap{\reff@jnl{A\&A}}              
\def\aapr{\reff@jnl{A\&A~Rev.}}        
\def\aaps{\reff@jnl{A\&AS}}            
\def\azh{\reff@jnl{AZh}}               
\def\baas{\reff@jnl{BAAS}}             
\def\jrasc{\reff@jnl{JRASC}}           
\def\memras{\reff@jnl{MmRAS}}          
\def\mnras{\reff@jnl{MNRAS}}           
\def\pra{\reff@jnl{Phys.Rev.A}}        
\def\prb{\reff@jnl{Phys.Rev.B}}        
\def\prc{\reff@jnl{Phys.Rev.C}}        
\def\prd{\reff@jnl{Phys.Rev.D}}        
\def\prl{\reff@jnl{Phys.Rev.Lett}}     
\def\pasp{\reff@jnl{PASP}}             
\def\pasj{\reff@jnl{PASJ}}             
\def\qjras{\reff@jnl{QJRAS}}           
\def\skytel{\reff@jnl{S\&T}}           
\def\solphys{\reff@jnl{Solar~Phys.}}   
\def\sovast{\reff@jnl{Soviet~Ast.}}    
\def\ssr{\reff@jnl{Space~Sci.Rev.}}    
\def\zap{\reff@jnl{ZAp}}               
\def\nat{\reff@jnl{Nature}}            

\title[Bayesian optimal reconstruction]{Bayesian optimal reconstruction of the primordial
 power spectrum}
  
\author[]
  {M.~Bridges\thanks{E-mail: m.bridges@mrao.cam.ac.uk}, F.~Feroz,
   M.P.~Hobson and A.N.~Lasenby\\
  Astrophysics Group,
      Cavendish Laboratory, JJ Thomson Avenue,
      Cambridge CB3 0HE, UK\\
}

\date{Accepted ---. Received ---; in original form \today}
\pagerange{\pageref{firstpage}--\pageref{lastpage}}
\pubyear{2007}

\begin{document}
\label{firstpage}
\maketitle

\begin{abstract}
 The form of the primordial power spectrum
 has the potential to differentiate strongly between competing models of perturbation generation in the
 early universe and so is of considerable importance. The recent release of five years of 
 WMAP observations have confirmed the general
picture of the primordial power spectrum as deviating slightly from scale invariance 
with a spectral tilt parameter of $n_{\rm s} \sim 0.96$. Nonetheless, many attempts have been made to 
isolate further features such as breaks and cutoffs using a variety of methods, some
employing more than $\sim 10$ varying parameters.
In this
paper we apply the robust technique of Bayesian model selection to reconstruct the
\emph{optimal} degree of structure in the spectrum. We model the spectrum simply and
generically as piecewise linear in $\ln$ $k$ between `nodes' in $k$-space whose amplitudes are allowed
to vary. The number of nodes and their $k$-space positions
are chosen by the Bayesian evidence so that we can identify both the complexity and location of 
any detected features. 
Our optimal reconstruction contains, perhaps, surprisingly few features, the data
preferring just three nodes. This reconstruction allows for a degree of scale 
dependence of the tilt with the `turn-over' scale occuring around $k \sim 0.016$ Mpc$^{-1}$.
More structure is penalised by the evidence as
over-fitting the data, so there is currently little point in attempting reconstructions that
are more complex.  
\end{abstract}


\section{Introduction}

The recent release by the Wilkinson Microwave Anisotropy Probe (WMAP) of five years of
observations have confirmed that the primordial spectrum of density perturbations is
consistent with being purely adiabatic and \emph{close} to scale invariant, in perfect
harmony with the simplest inflationary scenarios. This agreement appears remarkably robust 
when extended to independent datasets such as measures of the matter power spectrum from 
galaxy redshift surveys \citep{TegmarkII}. Alternative models of the spectrum containing
various features have been considered.  These include an
exponential large scale cutoff \citep{efstathioub} to explain the quadrupole power decrement,
 and theoretically motivated spectra to model the 
inflationary potential \citep{Contaldi2007} or account for discontinuities from 
early universe phase 
transitions \citep{Barriga}. 
Reconstructions of the spectrum, limiting \emph{a priori} assumptions about its 
structure,
have typically involved fitting some basis functions, such as wavelets \citep{Wang},
some deconvolution method (\citealt{Souradeep}; \citealt{Silk}) or
directly `binning' the spectrum into an arbitrary number of band powers \citep{Bridle}.
However, most previous methods fail to account for Occam's razor since they assume that more
complexity, and typically more `detected' features, are necessarily important in
explaining the data. Recently \citet{Verde07} reconstructed the spectrum, while minimising
the level of complexity needed via a cross-validation with a `hold-out' portion of the data. 
This approach is a timely progression, but in this paper we attempt a more statistically
robust procedure with an optimal
reconstruction using the Bayesian evidence to decide how much detail one should
fit and where it is located in $k$-space, based solely on the data.

\section{Parameterisation of the Primordial Spectrum}
Inflationary models generically predict the initial spectrum of scalar
density perturbations to be close to scale invariant with just slight
scale dependence, commonly called \emph{tilt}, a red (blue) 
tilt for decreasing (increasing) amplitude at smaller scales. 
Theoretical motivation for this
form is found in the slow-roll formulation of inflation. Previous studies 
(e.g. \citealt{Leach}
\& \citealt{Peiris}) have used spectral models defined explicitly by the physical slow-roll
parameters but here we define the spectrum essentially empirically using a spectral amplitude, 
$A_s$, a spectral index or tilt parameter $n_{\rm s}$ and a \emph{running} parameter $n_{\rm run} 
\equiv \frac{d n_{\rm s}}{d \ln k}$ denoting any tilt scale dependence: 
\begin{equation}
\mathcal{P}(k) = A_s \left( \frac{k}{k_0} \right)^{n_{\rm s} - 1 + \frac{1}{2} 
\ln \left( \frac{k}{k_0}\right) n_{\rm run}},
\label{equation:tilt}
\end{equation} 
where $k_0$ denotes the scale about which the tilted spectrum pivots which 
throughout we set at 
$0.05$ Mpc$^{-1}$. It has been shown previously \citep{Trotta} that this
parameterisation, although not physical in itself, does within suitable prior ranges 
adequately model the inflationary primordial spectrum. 

The parameterisation described by Eqn.~\ref{equation:tilt} encompasses the most commonly tested
power spectra, namely: the scale invariant or Harrison-Zel'dovich spectrum 
(in which $1 - n_{\rm s} =
n_{\rm run} = 0$), the tilted spectrum ($n_{\rm run} = 0$) and a running spectrum in which the tilt
becomes a function of scale ($n_{\rm run} \ne 0$). 
To these we can add a `cutoff' spectrum which allows
$\mathcal{P}(k)$ to drop to zero below some variable cutoff scale and above which behaves
like a tilted spectrum.
We shall use this as a
simple test as to whether the addition of some cutoff feature is actually required by the data.


In this paper, however, we are primarily interested in determining structure 
in the primordial spectrum using an optimal model-free reconstruction. We use the
Bayesian evidence as discriminator in fitting a simple spectrum based on linear interpolation
between a set of amplitude-varying \emph{nodes} in $k$-space. This is essentially the same
\emph{binning} format as that used previously by a number of authors
(\citealt{Bridle}, \citealt{Bridgesa}, \citealt{Bridgesc}, \citealt{Spergel2006}) however here 
we aim to allow the data to decide upon the location \emph{and} number of nodes via the
evidence.  

In the background cosmology we allow the possibility of a non-flat $\Lambda$CDM cosmology 
specified by the following five parameters:
the physical baryonic matter density $\Omega_b h^2$, the physical dark
matter density $\Omega_{\rm dm} h^2$, the ratio of the sound horizon to angular
diameter distance $\Theta$, the optical depth to reionisation $\tau$ and the curvature density
$\Omega_k$, where the corresponding priors are listed in Table \ref{table:priors}. 
Additionally we allow a contribution to the small-scale power in the CMB spectrum from
Sunyaev-Zeldovich fluctuations as performed in the WMAP analysis (\citealt{Dunkley},
\citealt{Komatsu}).

The structure of the paper is as follows: in section~\ref{section:multinest} we describe
basic model selection and our algorithm, in section~\ref{section:datasets} we list the
individual datasets and discuss the combinations used,
in section~\ref{section:H_Z} 
we will review the
current status of the standard, scale-invariant, tilted and running parameterisations of
the power spectrum in light of the WMAP5 data and test the possibility of a large-scale
cutoff. We then briefly discuss the
consistency of the datasets using a quantifiable Bayesian measure in
section~\ref{section:consistency}. The
remainder of the paper is then devoted to our optimal reconstruction 
(section~\ref{section:reconstruction}) and our conclusions
(section~\ref{section:conclusions}).  
 
\begin{table}
\begin{center}
\caption{Priors of the base cosmological parameters.}
\begin{tabular}{|c|}
    \hline
 $0.018 \leq \Omega_b h^2 \leq 0.032$\\
 $0.04 \leq \Omega_{dm} h^2 \leq 0.16$\\
 $0.98 \leq \Theta \leq 1.1$\\ 
 $0.01 \leq \tau \leq 0.5$\\
 $-0.1 \leq \Omega_k \leq 0.1$\\  
    \hline
\end{tabular}
\label{table:priors}
\end{center}
\end{table}

\section{Bayesian Inference}
\label{section:multinest}
The Bayesian methodology provides a logical and consistent approach to extracting
inferences from a set of data. Given a model, or hypothesis $H$ defined by a set of
parameters $\mathbf{\Theta}$, Bayes' theorem tell us how to determine the 
probability distribution of those parameters given the data $\mathbf{D}$:
\begin{equation} \Pr(\mathbf{\Theta}|\mathbf{D}, H) =
\frac{\Pr(\mathbf{D}|\,\mathbf{\Theta},H)\Pr(\mathbf{\Theta}|H)}
{\Pr(\mathbf{D}|H)},
\end{equation}
where for future simplicity we define $\Pr(\mathbf{\Theta}|\mathbf{D}, H) \equiv P(\mathbf{\Theta})$
as the posterior probability distribution of the parameters,
$\Pr(\mathbf{D}|\mathbf{\Theta}, H) \equiv
\mathcal{L}(\mathbf{\Theta})$ as the data likelihood, and
$\Pr(\mathbf{\Theta}|H) \equiv \pi(\mathbf{\Theta})$ as the prior.
Of particular importance here is the Bayesian evidence term 
$\Pr(\mathbf{D}|H) \equiv \mathcal{Z}$.

To obtain parameter constraints given a model the evidence is often 
ignored since it is independent of the parameters $\mathbf{\Theta}$. The posterior 
distribution is simply constructed by Monte Carlo sampling from the combined distribution
$P(\mathbf{\Theta}) \propto \mathcal{L}(\mathbf{\Theta}) \pi(\mathbf{\Theta})$. Typically 
most of the posterior \emph{weight} lies in a relatively small range of $\mathbf{\Theta}$ 
and so using some importance sampling procedure, like Metropolis-Hastings, one quickly
generates estimates of the best-fitting parameter values and their variances.  

Bayesian model selection also relies on the posterior distribution and is based on its
\emph{normalisation} over the parameter space $\mathbf{\Theta}$. This term is in fact given
by the evidence $\mathcal{Z}$ and can be computed by performing the integral:
\begin{equation}
\mathcal{Z} =
\int{\mathcal{L}(\mathbf{\Theta})\pi(\mathbf{\Theta})}d^N \mathbf{\Theta},
\label{eq:3}
\end{equation} 
where $N$ is the dimensionality of the parameter space. Thus $\mathcal{Z}$ can be 
defined as the
average of the likelihood over the prior.
The evidence naturally 
incorporates Occam's razor: a simpler theory with a more compact
parameter
space will have a larger evidence than a more complicated one, unless
the latter is significantly better at explaining the data.  

The question of which model best describes the data can then be addressed by 
comparing the properly normalised posterior probability distributions calculated for two 
hypotheses $H_0$ and $H_1$.
\begin{equation}
\frac{\Pr(H_1|\mathbf{D})}{\Pr(H_0|\mathbf{D})}
=\frac{\Pr(\mathbf{D}|H_1)\Pr(H_1)}{\Pr(\mathbf{D}|
H_0)\Pr(H_0)}
=\frac{\mathcal{Z}_1}{\mathcal{Z}_0}\frac{\Pr(H_1)}{\Pr(H_0)},
\label{eq:3.1}
\end{equation}
where $\Pr(H_1)/\Pr(H_0)$ is the a priori probability ratio for the
two models, which can be set to unity if we have no reason to prefer hypothesis $H_0$
over $H_1$ initially. For convenience the ratio of evidences $\mathcal{Z}_1/\mathcal{Z}_0$
(or equivalently the difference in log evidences $\ln \mathcal{Z}_1 - \ln \mathcal{Z}_0$)
 is
often termed the \emph{Bayes' factor} $\mathcal{B}_{01}$. Interpreting the level of
significance one should ascribe to a given $\mathcal{B}$ value is often a matter of experienced
judgement, however a suitable guideline scale has been laid out by \citet{Jeffreys}. If
$ \mathcal{B} < 1 $ $H_1$ should not be favoured over $H_0$, $ 1 < \mathcal{B} <
2.5$ is significant,  
$ 2.5 < \mathcal{B} < 5$ is strong evidence while 
$\mathcal{B} > 5$ would be considered decisive. 

Computation of the multidimensional integral Eqn.~\ref{eq:3} is not a trivial 
task and approaches such as thermodynamic integration have previously been shown 
to be both slow and inaccurate. In this analysis we apply the method of nested sampling
\citep{Skilling} which transforms the $N$-dimensional integral in Eqn.~\ref{eq:3} to one dimension 
and computes it by drawing uniform samples from ever decreasing nested \emph{shells} in the
prior parameter space. We apply an algorithm based on this procedure
called {\sc MultiNest} which constrains the nested shells in the prior space 
with $N$-dimensional
ellipsoids (\citealt{Feroz2008a}; \citealt{Feroz2008b}). This approach results in an order of magnitude improvement in
efficiency  and accuracy over previous methods. 

\section{Datasets considered}
\label{section:datasets}
In this analysis we have divided the data into two categories: CMB only and CMB plus
observations of the matter power spectra from Large Scale Structure (LSS) surveys. This is
primarily designed so that we can test consistency across the datasets in an initial analysis
before carrying over a final set of data to our power spectrum reconstruction. We consider a number of CMB
experiments including the latest five year release from WMAP \citep{Hinshaw08} plus recent
results from the Arcminute Cosmology Bolometer Array [ACBAR; \citealt{ACBAR2008}] which
should be uniquely useful here due to their tight constraints out to small angular scales.
In addition we include Cosmic Background Imager observations [CBI; \citealt{CBII};
\citealt{CBIII}] and Balloon Observations of Millimetric Extra-galactic Radiation and
Geophysics [{\sc BOOMERanG}; \citealt{BOOMI}; \citealt{BOOMII}; \citealt{BOOMIII}]. 
LSS data includes the luminous red galaxy (LRG) subset D4 of the Sloan 
Digital Sky Survey [SDSS;
\citealt{SDSSII}] and the two degree field survey [2dF; \citealt{Cole}]. 
We allow for
modelling of non-linearities and galaxy biasing of the matter power spectrum
in the LRG sample using the transfer function
defined by \citet{Cole} \footnote{$P_{\rm Non-linear}(k) = b^2 \frac{1+Qk^2}{1+Ak} 
P_{\rm Linear}(k)$}.
We analytically marginalise over the parameter combination $Qb^2$ and set $A=1.4$, as 
shown by \citet{Cole} to be adequate.

\section{Simple Power Spectrum Models}
\label{section:H_Z}
Many previous analyses have considered the four most basic parameterisations described in 
Section 2 in light of WMAP
 observations plus a plethora of higher resolution CMB and Large
Scale Structure (LSS) data. Here we will briefly summarise the current status of these models.
The first year WMAP [WMAP1] data on its own had no preference for a tilt  ($n_{\rm s} =
0.99\pm 0.04$) but the inclusion of higher resolution CMB data and LSS data induced a marked
red-tilt \citep{Spergel2003}. By year three of WMAP [WMAP3], with tighter constraints on the
second and third acoustic peaks, a red tilt became discernible even without additional
datasets \citep{Spergel2006} $n_{\rm s} = 0.958 \pm 0.016$. The recent WMAP five
year release confirms the value at $\sim 0.96$ with a mean estimate of $0.963 \pm 0.015$
\citep{Komatsu}.
The position of a running spectral index has been more
controversial: WMAP1 alone preferred a large mean value of $n_{\rm run}$ though with 
little statistical significance,  
with WMAP3 alone however a value of $n_{\rm run}$ was found, that within
1$\sigma$ limits was deviant from zero. 
A number of authors (\citealt{Viel}; \citealt{Seljak}; \citealt{Bridgesc}) 
have subsequently found that in the case of WMAP3, running was almost completely removed on addition of the SDSS 
Ly-$\alpha$ 
forest data \citep{McDonaldI}. 
Ly-$\alpha$ data probes scales
($\sim$ Mpc), small in comparison to other datasets used, and so provides a long `lever arm'
for primordial spectrum analyses. However further discrepancies in other cosmological
parameters, at the level of almost 2$\sigma$, has cast some doubt on the conclusions made when
using this data so we do not include it here. 

Theoretically motivated priors on the tilt are easily extracted from the
slow-roll inflationary framework as $n_{\rm s} = 1 - 6 \epsilon + 2 \eta$, where $\epsilon$ and
$\eta$ are the slow-roll parameters. For the slow-roll conditions to be met 
we then require that 
$\epsilon \sim 0$ and that $\eta \ll 1$. If we assume that $\eta$ must be $\leq 0.1$ we
get $n_{\rm s} = 1 \pm 0.2$ \citep{Trotta}. Spectral running is expected to be small, in
fact $n_{\rm run}$ even at the level of $0.05$ would rule out all simple inflationary scenarios
\citep{Easther}. Thus if assuming slow-roll inflation we are free to set quite a 
tight prior
$-0.2 \leq n_{\rm run} \leq 0.2$. Uniform prior distributions over these ranges were adopted throughout. 

Using {\sc MultiNest} as described in section~\ref{section:multinest} a set of posterior
samples and model evidences were computed using the two basic datasets described in 
section~\ref{section:datasets} for the basic suite of models: H-Z spectrum, a tilted spectrum, a tilted spectrum
with running and a tilted spectrum with a large scale cutoff.
For now this simply serves as a useful sanity check for consistency 
between datasets, but later, in section~\ref{section:consistency} the appropriate Bayesian 
consistency measure will be applied to quantify any discrepancy.

\begin{figure}
\begin{center}
\psfrag{xlabel}{$n_{\rm s}$}
\includegraphics[width=0.6\linewidth, angle = -90]{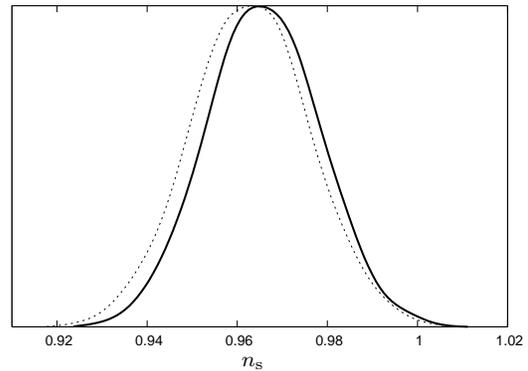}
\caption{Marginalised posterior probability of the spectral tilt $n_{\rm s}$ using CMB plus 
LSS data (solid) and CMB data alone (dotted).
Note: in this and all subsequent figures each posterior is normalised
independently.}
\label{figure:n_s}
\end{center}
\end{figure}

We will now discuss the most common set of parameters that are typcially used to
describe the primordial spectrum: $n_s$ from the tilted spectrum and 
$n_{\rm run}$ from the
tilted spectrum with running.
Figure~\ref{figure:n_s} shows the marginalised posterior
distribution on $n_{\rm s}$ from the tilted power spectrum
using CMB data alone and in a joint analysis with LSS data. We find a mean value
of $n_{\rm s} = 0.962 \pm 0.018$, this value shifting upwards only marginally when including LSS
($n_{\rm s} = 0.967$). These results are in good agreement with \citet{Komatsu} despite our relaxation of the requirement for universal flatness. Deviations
from $n_{\rm s} \approx 1$ such as these, at $\sim 2\sigma$ are now seen as persuasive 
evidence for a red-tilt. The \emph{Bayesian} evidence however would need a significantly
larger deviation (in fact closer to the level of $5\sigma$!) to conclude decisively that 
tilt was present. At present these results produce a Bayes' factor of $\mathcal{B}_{H-Z, n_{\rm s}} \sim
1.1 - 1.6$ (see Table~\ref{table:h_z_tilt_running_cutoff}), that is significant but certainly 
not strong evidence in favour of a tilt. 
Running in the spectrum remains ambiguous with CMB data alone (roughly a $1\sigma$ deviation
from $n_{\rm run} = 0$), but the addition of LSS data shifts the mean value to within $\pm
0.02$ of zero (Fig.~2). This effect has been observed on a number of
occasions (e.g. \citealt{Tegmark}, \citealt{Bridgesc}) and is due mainly to the excellent high-$k$
constraints coming from the LRG data. The evidence does not favour running in either dataset,
with $|\mathcal{B}_{H-Z, n_{\rm run}} \sim 0.4|$, just outside our estimated margin of error.
\begin{center}
\begin{figure}
\psfrag{xlabel}{$n_{\rm run}$}
\begin{centering}
\includegraphics[width=0.6\linewidth, angle = -90]{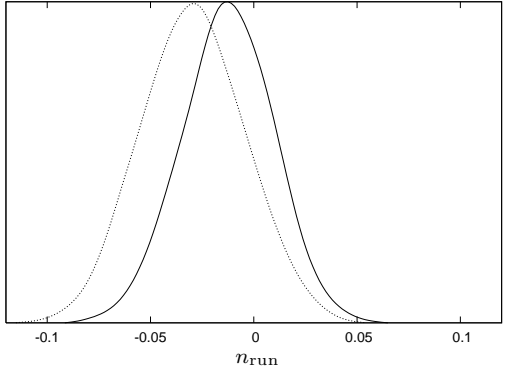}
\caption{Marginalised posterior probability of spectral running $n_{\rm run}$ using CMB plus 
LSS data (solid) and CMB data alone (dotted).}
\end{centering}
\label{figure:n_run}
\end{figure}
\end{center}
\begin{table}
\begin{center}
\caption{Bayes' factors comparing a scale invariant (H-Z) spectrum with models containing
tilt, running and a large scale cutoff using
both CMB alone and CMB + LSS data.}
\begin{tabular}{|c||c||c|}
    \hline
 \textbf{Model} &  CMB & CMB + LSS \\
    \hline
 H-Z	&  $0.0 \pm 0.3$ & $0.0 \pm 0.3$\\
 $n_{\rm s}$  &  $+1.6 \pm 0.3$ & $+1.1 \pm 0.3$\\
 $n_{\rm run}$ & $+0.4 \pm 0.3$ & $-0.4 \pm 0.3$\\
   $k_c$  & $+1.5 \pm 0.3$ & $+1.3 \pm 0.3$\\
    \hline
\end{tabular}
\label{table:h_z_tilt_running_cutoff}
\end{center}
\end{table}

\begin{figure}
\begin{center}
\psfrag{ylabel}{$\ell ( \ell + 1) C_{\ell} / 2\pi$ [$\mu K ^2$]}
\psfrag{xlabel}{$\ell$}
\includegraphics[width=0.6\linewidth, angle = -90]{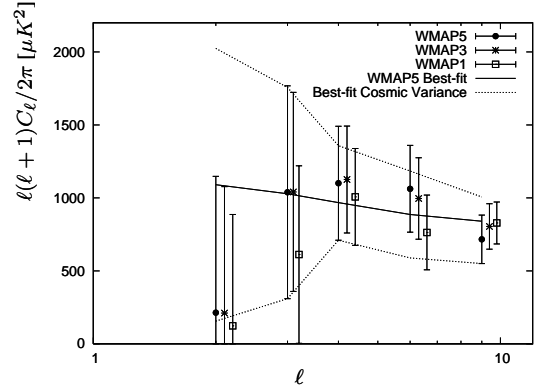}
\caption{Low-$\ell$ multipoles and $1\sigma$ error bars from three releases of WMAP data the 
best-fit fiducial power
spectrum based on WMAP5 inferences is also plotted and shows the associated cosmic variance limits. [Note
$\ell$ values are slightly offset for clarity.]}
\label{figure:low_l}
\end{center}
\end{figure}

Figure~\ref{figure:low_l} shows the measured $C_{\ell}$ values at low-$\ell$ for WMAP1, 3 and 5 with the
best-fit theoretical model (and corresponding cosmic variance limits) as determined by
\citet{Dunkley}. The mean $C_{\ell}$ estimators at both the quadrupole and octopole in WMAP1
are seen to be deviant from the fiducial model by close to the cosmic variance limit. The
situation changed somewhat in the three-year (and subsequently five-year) release so that 
now the octopole has shifted upwards to lie comfortably close to its expected value, but
the quadrupole remains anomalously low. The statistical significance has been questioned by
many authors (e.g. \citealt{efstathioua}) and spurious alignments between the affected
multipoles have been suggested as evidence of some large scale foreground contamination
\citep{Oliveira-Costa}. However here we shall assume that the effect is a real one and
attempt to explain the large-scale CMB decrement with a feature in the primordial spectrum. 

\begin{figure}
\begin{center}
\psfrag{xlabel}{$k_c$}
\includegraphics[width=0.6\linewidth, angle = -90]{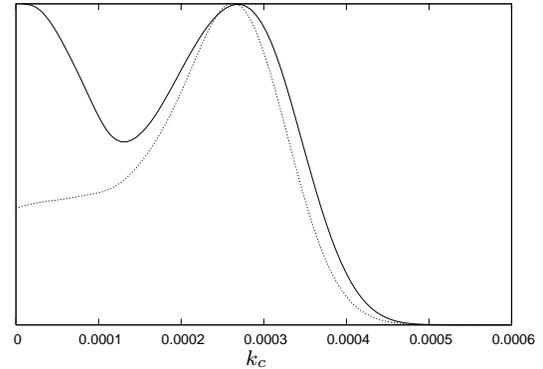}
\caption{Marginalised posterior probability of the large scale spectral cutoff $k_c$ 
using CMB plus 
LSS data (solid) and CMB data alone (dotted).}
\label{figure:k_c}
\end{center}
\end{figure}

Naturally, at present the data will prefer a model that includes a large scale cutoff, but
does the data find one \emph{necessary}? We can test this with a simple `cartoon' model
by abruptly curtailing 
a tilted spectrum below some variable scale $k_c$ so that its form is given by:
\begin{equation}\mathcal{P}(k) = \left\{ \begin{array}{ll}
         0,& \mbox{$k < k_c$}\\
     A_s \left(\frac{k}{k_0} \right)^{n_{\rm s}-1},& \mbox{$k \geq k_c$}
     \end{array}\right.
    \label{equation:cutoff}
\end{equation}
The marginalised posterior distributions for $k_c$ in Fig.~\ref{figure:k_c} show
a preferred scale around $2.7 \times 10^{-4}$ Mpc$^{-1}$, consistent with an angular
scale around $\ell = 2 - 4$ as expected. Interestingly although blind to scales around the 
cutoff, a joint analyses with LSS data shows a pronounced 
peak at $k_c \approx 0$ suggesting that the constraining power of, particularly LRG data, 
now matches current CMB data. In other words, now that constraints at smaller scales are 
becoming tighter, anomalies such as the cutoff are becoming less
important. The evidence confirms this (see Table~\ref{table:h_z_tilt_running_cutoff}) 
showing that the extra parameter is superfluous.  

The current position of these standard parameterisations then appears straightforward, with
CMB data alone and in joint analysis with LSS, a purely scale-invariant spectrum is
significantly disfavoured by the data. However the addition of a running parameter 
remains of dubious necessity with CMB data alone and is actually disfavoured when LSS
constraints are included. A large scale cutoff in the primordial spectrum remains a suitable
explanation of the WMAP quadrupole decrement but according to the evidence there is currently
no need to include it in the model. 

\section{Dataset Consistency}
\label{section:consistency}
Combining multiple datasets in joint analyses, in particular the recent inclusion of observations of
the baryonic acoustic oscillations in LSS surveys with CMB observations, have led to tight
constraints on the cosmological parameters \citep{TegmarkII}. Authors regularly comment on the
relative consistency between datasets by comparing the parameter constraints made with each
set individually and when combined, however little effort is normally made to quantify this consistency.
\citet{Marshall2006} established just such a method using the Bayesian evidence (see also
\citealt{Hobson2002}). 
This is important for our reconstruction as
experimental features, such as discontinuities on scales where observations meet may result
in false detections of spectral structure. The two datasets chosen, CMB and LSS, now
overlap considerably on scales starting around $k \sim 0.02$ Mpc$^{-1}$. If a data 
inconsistency were to exist it would likely appear as a feature close to this scale. 
Curiously such a feature has been identified, \citet{Verde} detected a deviation 
from a simple tilt around $k \sim 0.01$ Mpc$^{-1}$. This effect was
strongest when using WMAP data alone, appearing considerably reduced in joint analyses with
other higher resolution CMB and LSS data. 
Here we
will apply a Bayesian consistency check to assess whether we can be justified in combining
these datasets in our analyses.

Consider the null hypothesis $H_0$ that given two independent sets of data there is one model
and one set of parameters to explain them. In this case we would say that the datasets are
`consistent'. However we would really like a quantitative measure by which to assess
this consistency. If we consider the alternative, $H_1$, that each dataset separately prefers a
different set of parameters, we can then construct the Bayes' factor between the two
hypotheses as:
\begin{center}
\begin{eqnarray}
\mathcal{B}_{01} &\equiv& \frac{\Pr(\mathbf{D}|H_0)}{\Pr(\mathbf{D}|H_1)}\\
		&=& \frac{\mathcal{Z}_0(\mathbf{D})}{\prod_i \mathcal{Z}_1(\mathbf{D}_i)}
\end{eqnarray}
\end{center}   
where we have written $\Pr(\mathbf{D}|H_1)$ as the product of evidences from each individual 
(independent) dataset $D_i$. In this form consistency can easily be checked by computing the
joint evidence and the evidence due to each dataset separately. As with any other hypothesis
test we can assess the appropriate model with the aid of the Jeffreys' scale based on
the final Bayes' factor.

Table~\ref{table:consistency} lists the appropriate Bayes' factors for each model based on
our two datsets: CMB and CMB+LSS. Firstly, all factors are positive and greater
than unity, confirming that these sets of data are indeed all essentially free from
discrepancies. On the Jeffreys' scale, hypothesis $H_0$ that the datasets are consistent, is
favoured \emph{significantly}.
The highest degree of consistency occurs for the H-Z model, this is
not surprising as both datasets provide equivalent constraints on the amplitude of 
fluctuations. 
Where we did observe differences in parameter constraints with 
the running and cutoff models, we can see 
how this measure has quantified the discrepancy. For instance, the addition of LSS data,
led to slightly tighter constraints on the parameter $n_{\rm run}$
(as well as being pulled closer to zero) (Fig.~\ref{figure:n_run}) and this difference has
lowered the evidence in favour of consistency from nearly two log units to
$\sim 1$. A similar but less pronounced effect is observed with the cutoff model. 

The deviations seen are minor. The worst discrepancy found, using the running model,
was still consistent with CMB data, with odds of around 3:1 in favour (i.e. $e^{\Delta \ln \mathcal{Z}} =
e^{\mathcal{B}_{01}}$) while under
the assumption of scale invariance the datasets are consistent at around 14:1 in favour. 
These differences are best explained by the superior small scale constraints that are
possible when using LSS
data rather than a genuine inconsistency, and we feel it is justified to perform our
reconstruction using the joint set
of data given the increased constraining power possible.

\begin{table}
\begin{center}
\caption{Bayes' factors comparing the assumption of dataset consistency 
($H_0 =$ consistent, $H_1 =$ inconsistent) using CMB + LSS
datasets for each of the models considered above.}
\begin{tabular}{|c||c|}
    \hline
 \textbf{Model} & $\mathcal{B}_{01}$ \\
    \hline
 H-Z	&  $+2.6 \pm 0.3$\\
 $n_{\rm s}$  &  $+1.9 \pm 0.3$\\
 $n_{\rm run}$ & $+1.1 \pm 0.3$\\
   $k_c$  & $+1.5 \pm 0.3$\\
    \hline
\end{tabular}
\label{table:consistency}
\end{center}
\end{table}

\section{Optimal Power Spectrum Reconstruction}
\label{section:reconstruction}
The degree of structure that can or should be usefully constrained in the
primordial spectrum has been a source of increasing debate in the literature. 
Recently \citet{Verde07} applied a smoothing spline technique \citep{Sealfon05}
that attempts, via cross-validation with part of the data, to minimise the
complexity of the parameterisation. This approach selects an initial set of `knots'
that are fixed in $k$ space but whose amplitudes may vary, and through which various splines are fitted, thus
constructing the primordial spectrum. This approach will preferentially identify smooth structures
rather than sharp breaks, and while it is true that most deviations from scale invariance
given the slow-roll assumption will be smooth, we do not believe the data is currently
accurate enough for this to be the limiting factor for an analysis.
We have thus attempted to use the simplest
reconstruction possible, while still maintaining continuity, by linearly interpolating between
a set of \emph{nodes}, at which we allow the amplitude to vary. Our reconstructions gain
complexity by the addition of new nodes and on estimating the evidence for each
reconstruction one can decide exactly
the level of parameterisation deemed necessary by the data.

We start with one node, 
see Fig.~5 (a), so our base model is equivalent to the scale-invariant H-Z spectrum.
The next model, (b) allows for two, sufficiently separated, independently varying nodes, thus
emulating a tilted spectrum. We then add a third node (c), spaced logarithmically midway between
two existing nodes. This process continues, at each
stage the additional node being added between the existing ones, so that at the fourth stage
there are two possibilities, (d) and (e). At the fifth stage there are three possibilities, at the sixth, four
and so on. One can see that by such a process, using the evidence as the model discriminator 
at each stage, not
only are the number of parameters constrained but also the location of features in $k$-space, so
that we can faithfully reconstruct \emph{both} the degree and position of any spectral
structure. It should also be clear that if we \emph{branch} at one reconstruction by
accepting a new node at some position (say the lower $k$ node in (d) rather than (e)), 
we still retain the option of splitting the unaccepted region later (i.e. in (h)). 
Thus we fully explore the options in feature space and should hierarchicly detect as much
structure as the data will allow. 

The only assumptions required are the positions of the two extremal nodes,
$k_{\rm min}$ and $k_{\rm max}$. These bounds were chosen to lie at sufficiently large ($k_{\rm max} =
2.7$ Mpc$^{-1}$) and small ($k_{\rm min} = 0.0001$ Mpc$^{-1}$) scales so as safely to
encompass all
current observational probes \emph{and} crucially, when more than 2 nodes are used, to allow
the spectrum to tend \emph{naturally} to zero power, particularly on small scales.
A conservative amplitude prior of $0$-$55$ $\times 10^{-10}$ was used throughout on all nodes.

\begin{figure*}
\psfrag{xlabel}{\tiny $k$}
\psfrag{ylabel}{\tiny $\mathcal{P}(k) \times 10^{-10}$}
\centering
\subfigure[{\bf 1:} $\mathcal{B}_{11} = 0.00 \pm 0.30$]{\includegraphics[width = 0.17\linewidth,angle =-90]{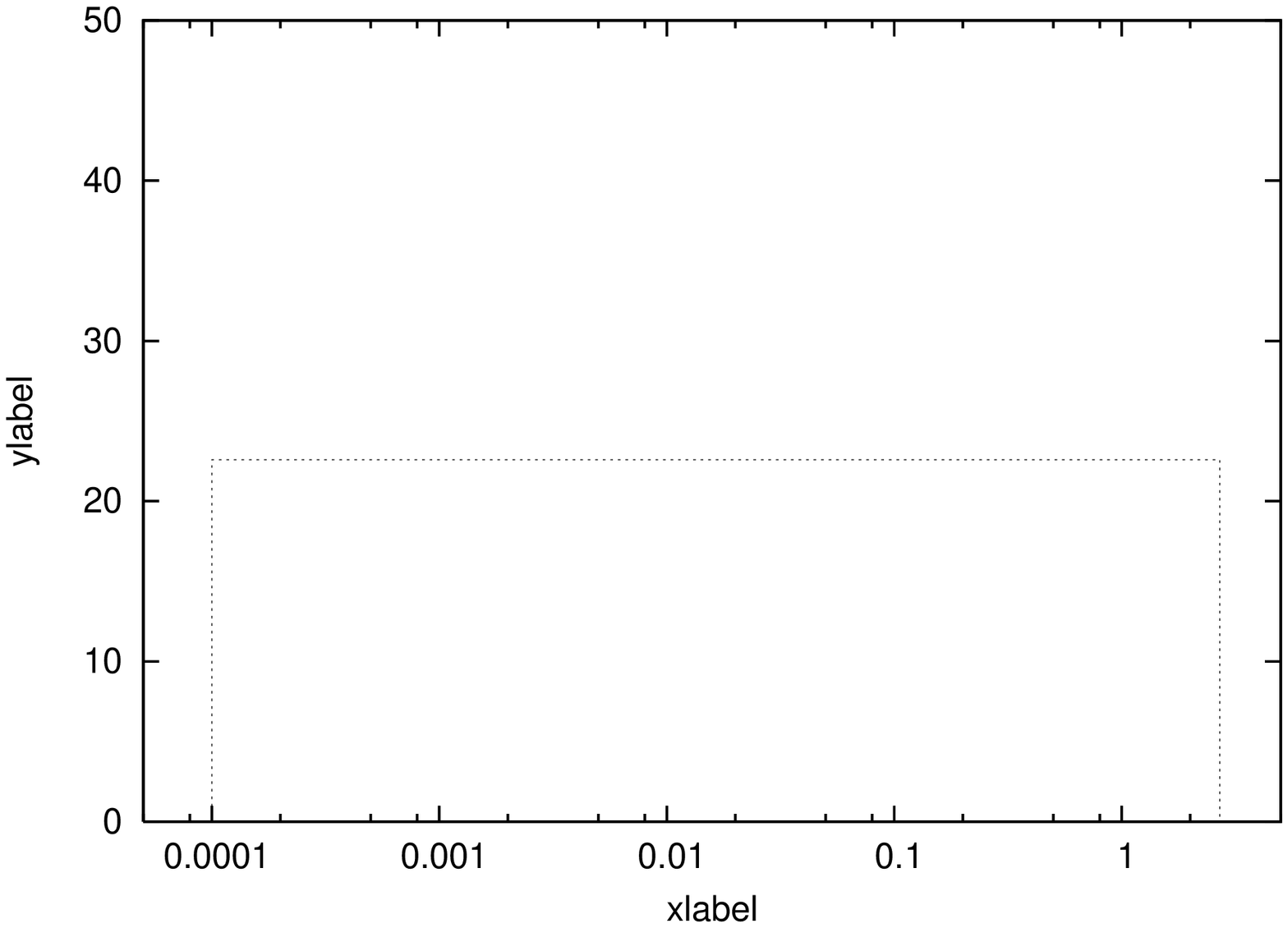}}\\
\subfigure[{\bf 2:} $\mathcal{B}_{21} = +0.66 \pm 0.30$]{\includegraphics[width = 0.17\linewidth, angle =-90]{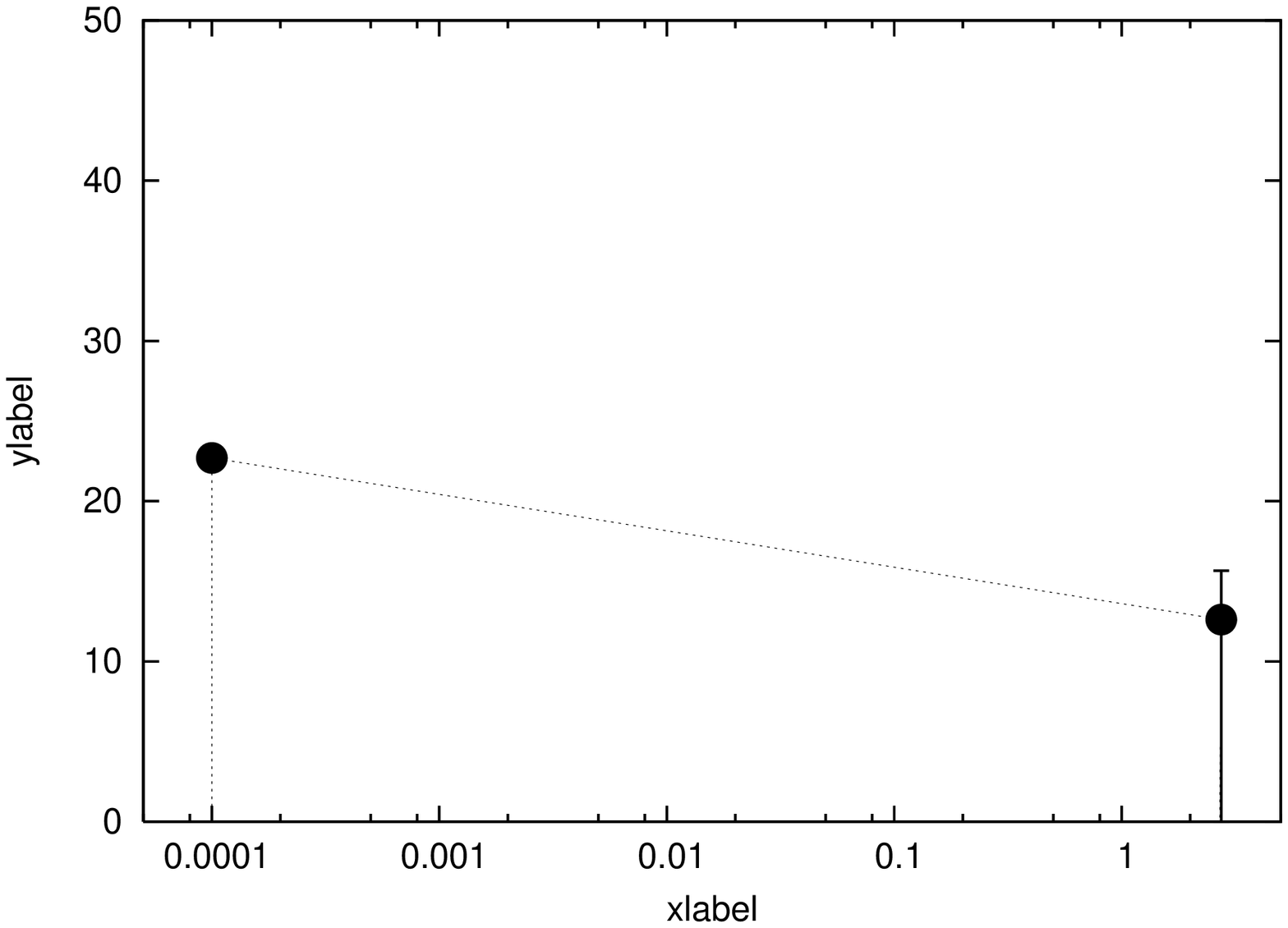}}\\
\subfigure[$\mathbf{3:}$ $\mathcal{B}_{31} = +1.08\pm 0.30$]{\includegraphics[width = 0.17\linewidth, angle =-90]{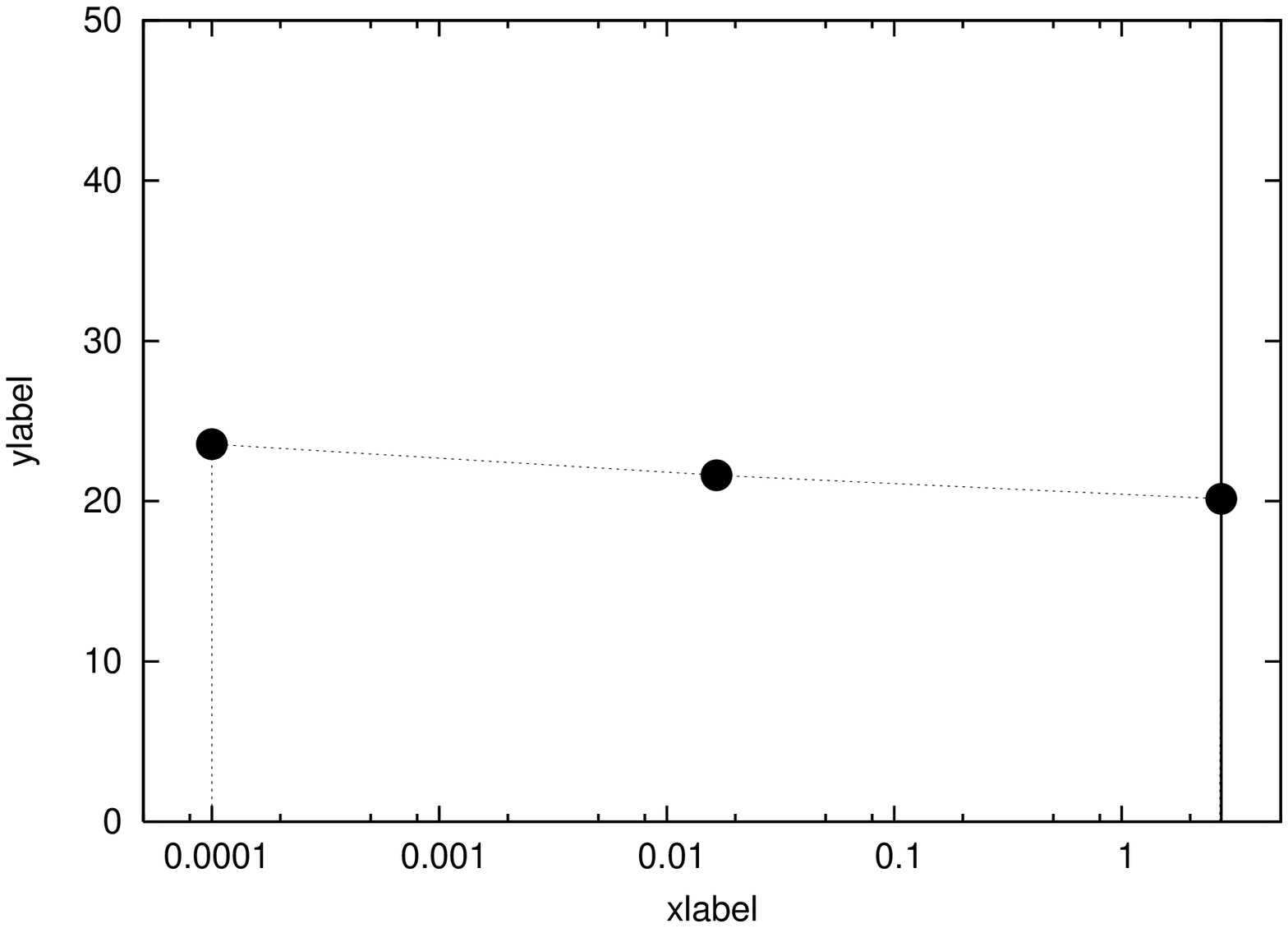}}\\
\subfigure[$\mathbf{4_{\rm I}:}$ $\mathcal{B}_{4_{\rm I}1} = -0.34 \pm 0.30$]{\includegraphics[width = 0.17\linewidth, angle=-90]{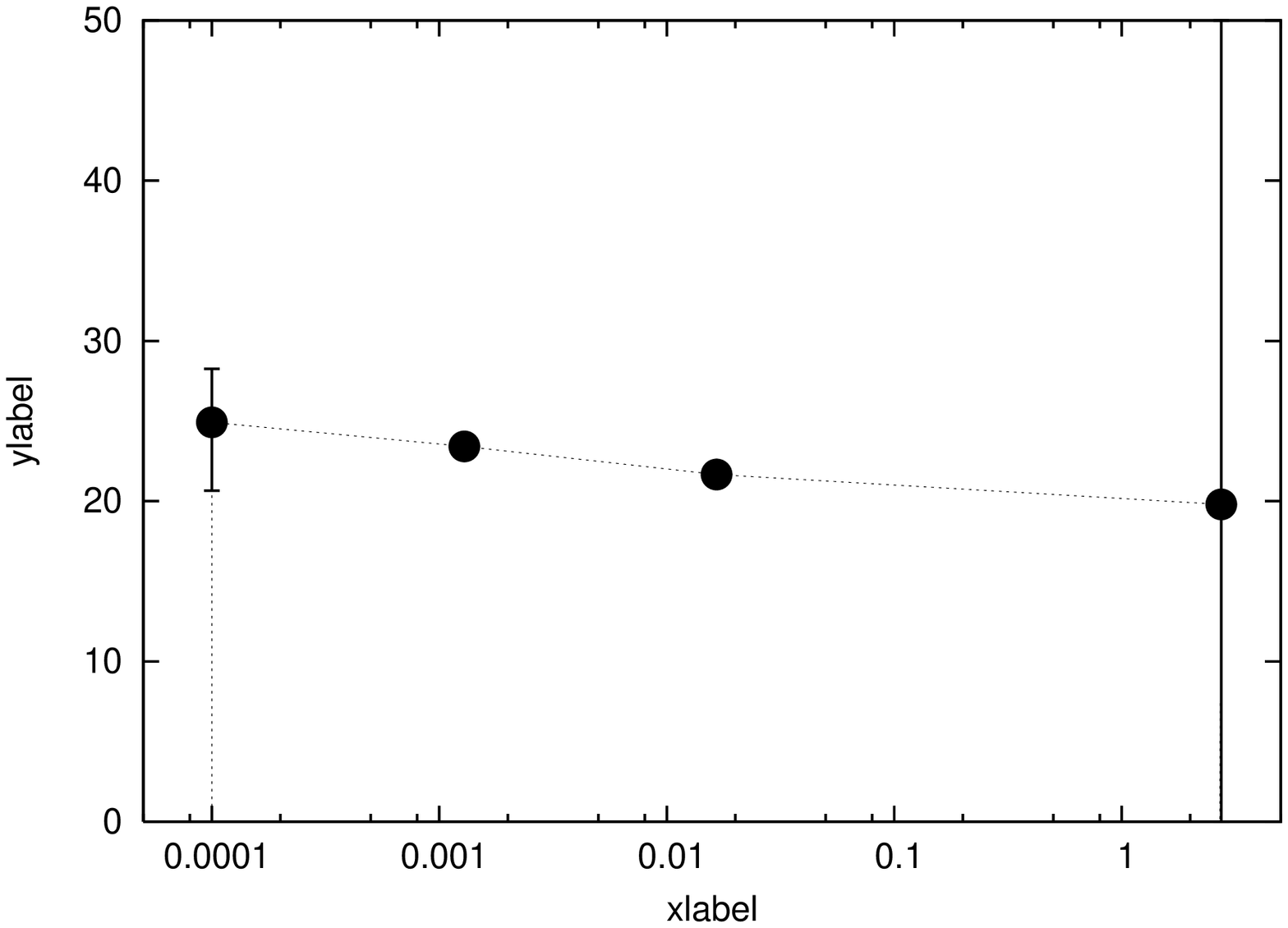}}
\subfigure[$\mathbf{4_{\rm II}:}$ $\mathcal{B}_{4_{\rm II}1} = -1.41 \pm 0.30$]{\includegraphics[width = 0.17\linewidth, angle =-90]{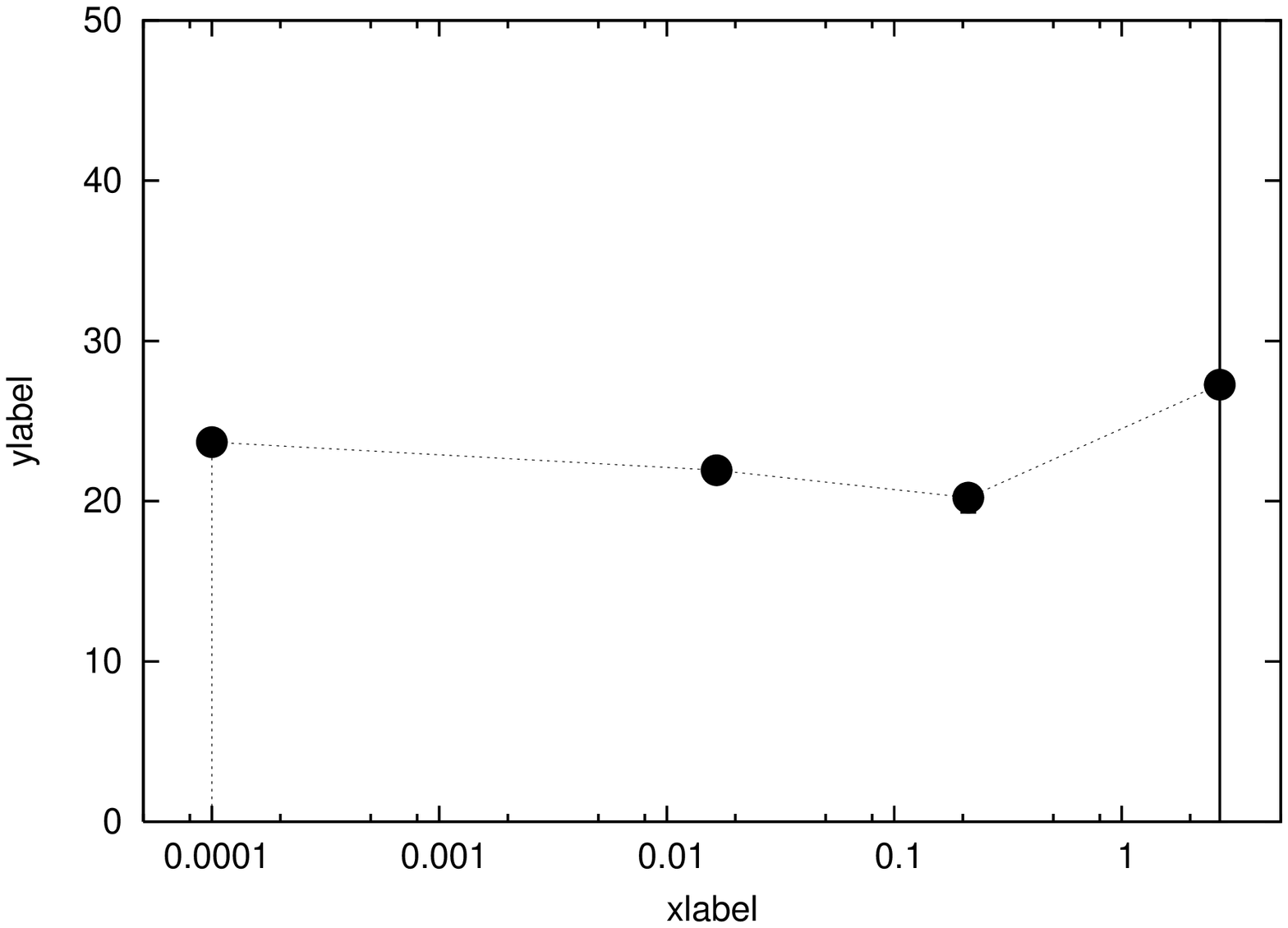}}\\

\subfigure[$\mathbf{5_{\rm I}:}$ $\mathcal{B}_{5_{\rm I}1} = -0.51 \pm 0.30$]{\includegraphics[width = 0.17\linewidth, angle=-90]{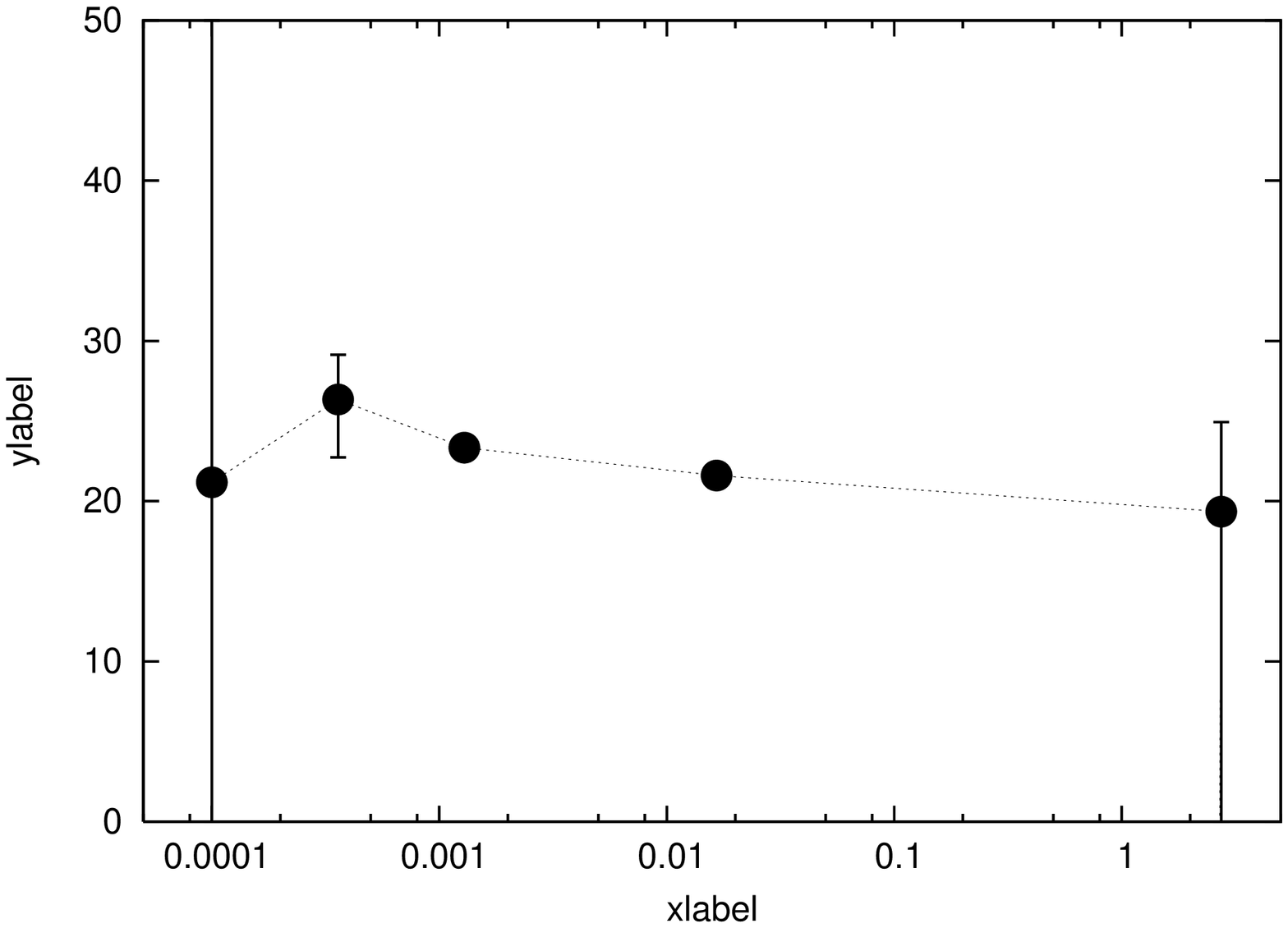}}
\subfigure[$\mathbf{5_{\rm II}:}$ $\mathcal{B}_{5_{\rm II}1} = -2.41 \pm 0.30$]{\includegraphics[width = 0.17\linewidth,angle=-90]{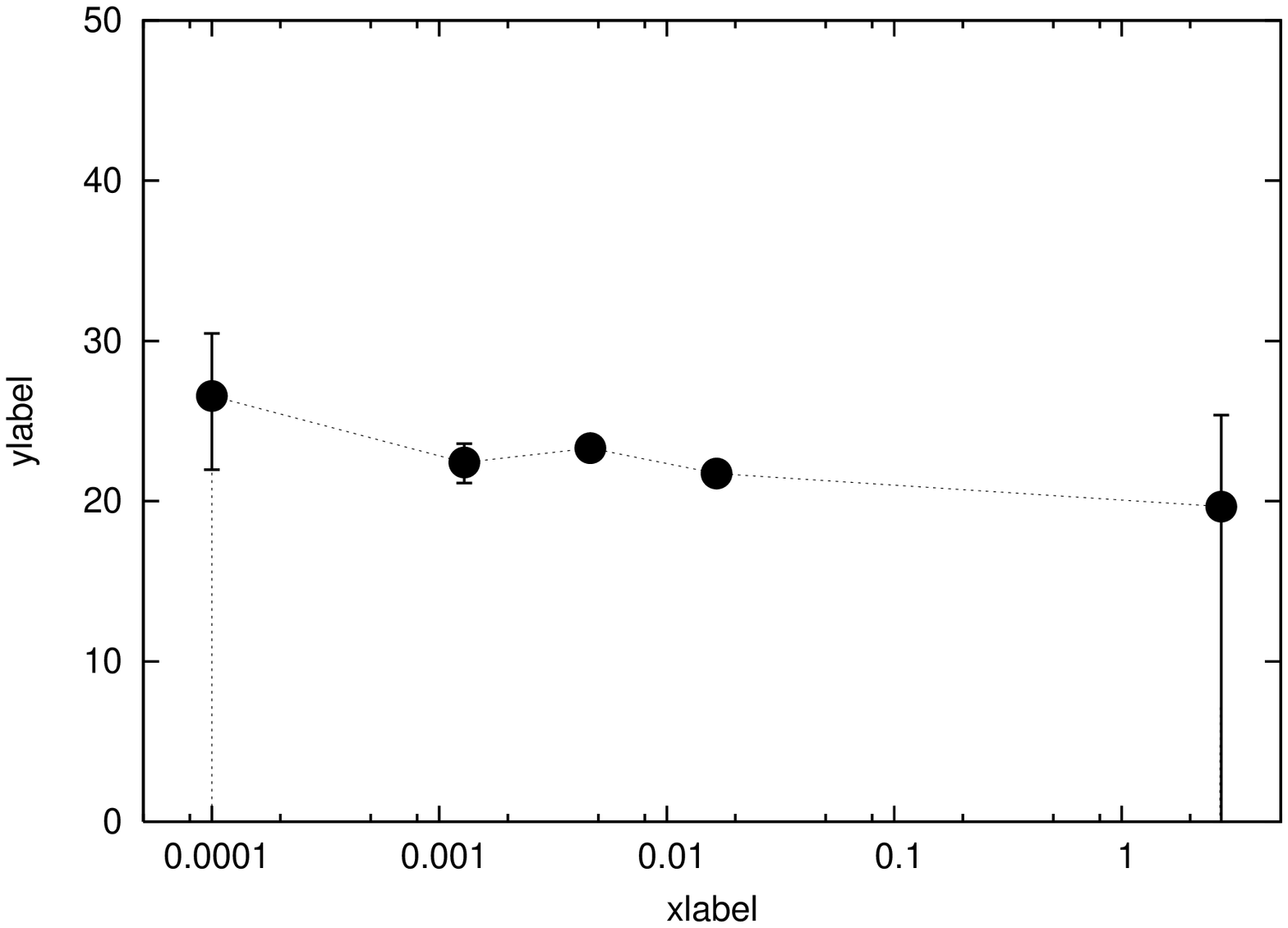}}
\subfigure[$\mathbf{5_{\rm III}:}$ $\mathcal{B}_{5_{\rm III}1} = -2.05 \pm 0.30$]{\includegraphics[width = 0.17\linewidth, angle =-90]{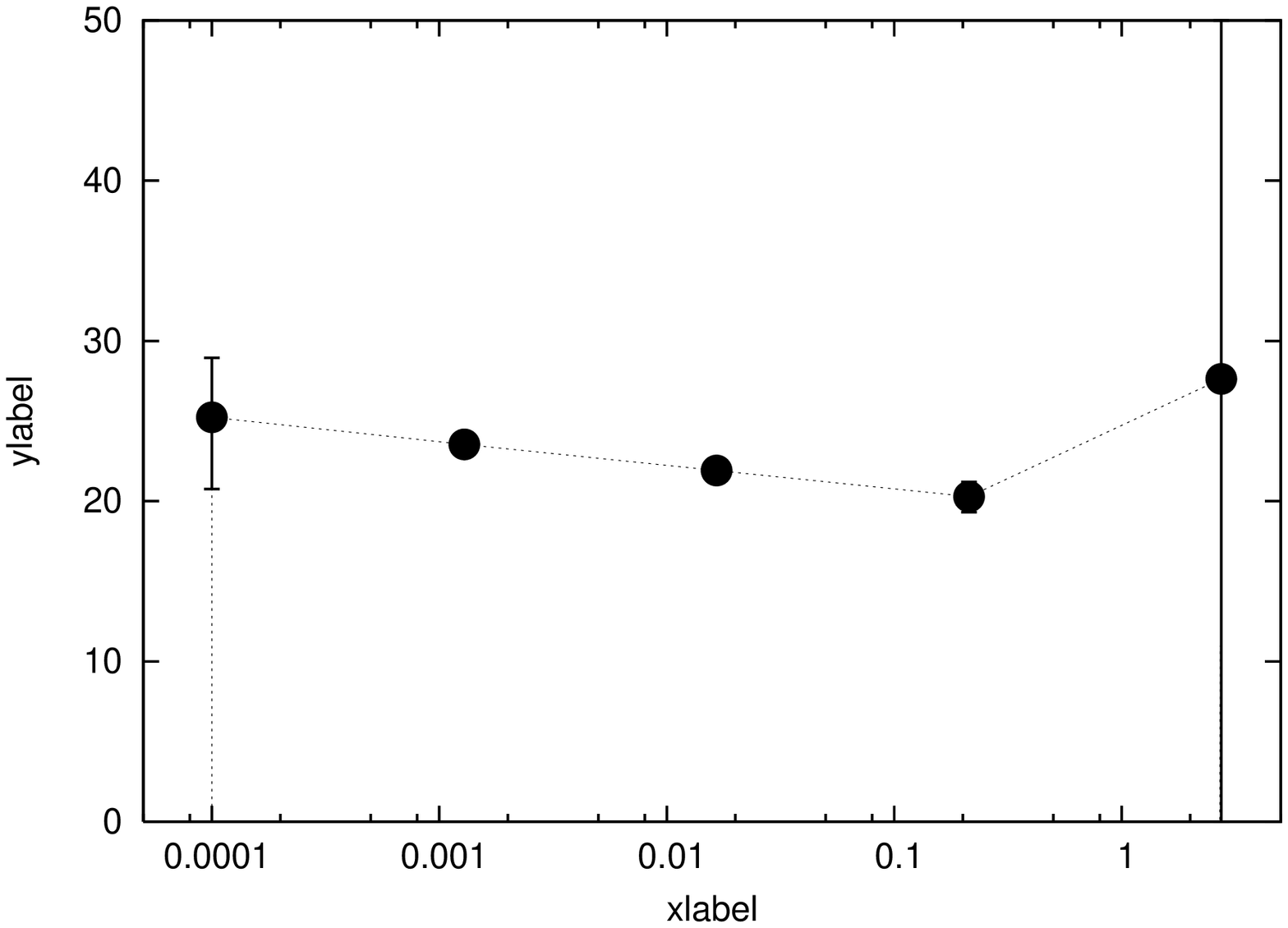}}\\

\subfigure[$\mathbf{6_{\rm I}:}$ $\mathcal{B}_{6_{\rm I}1} = -0.21 \pm 0.30$]{\includegraphics[width = 0.17\linewidth, angle=-90]{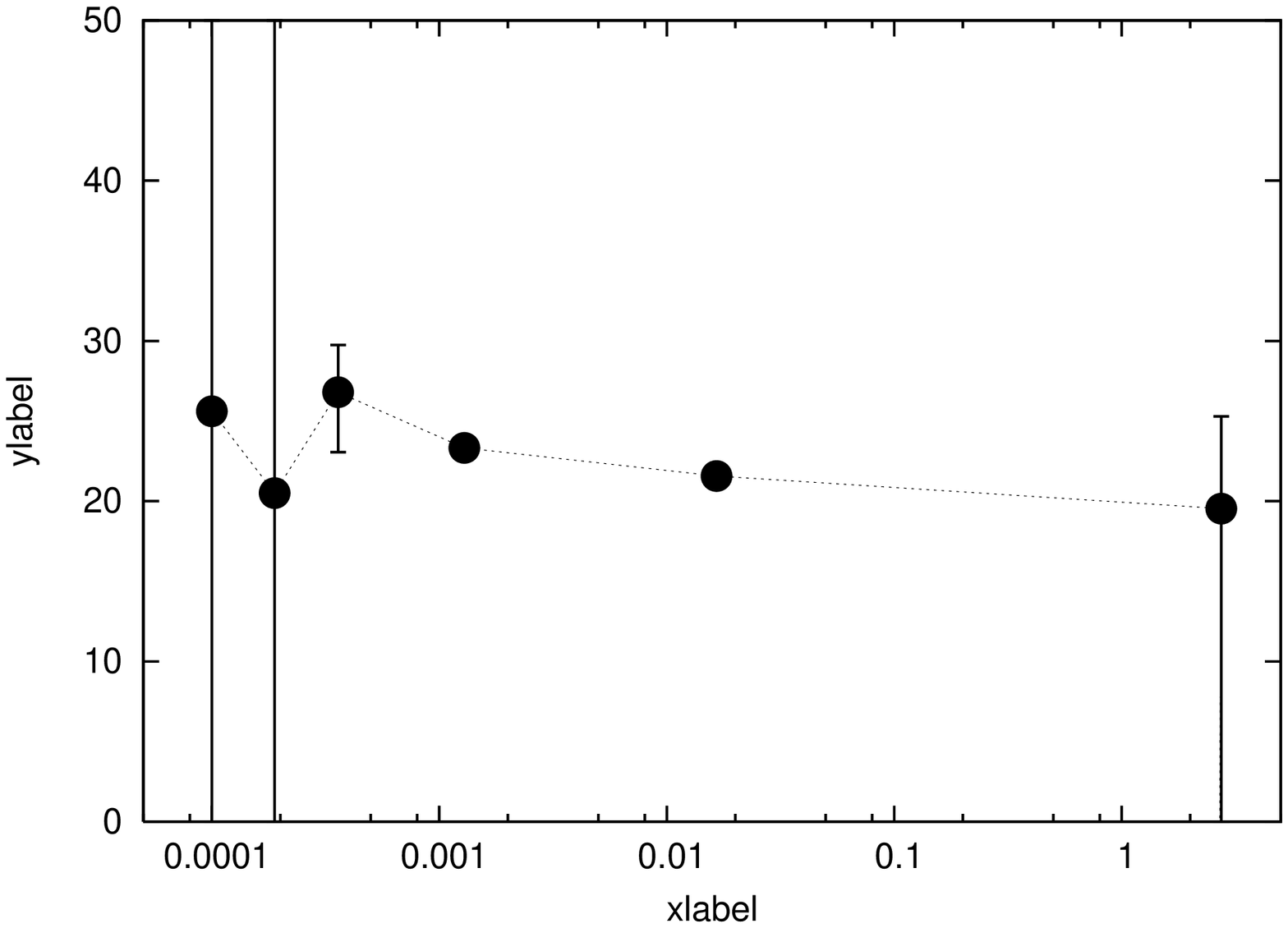}}
\subfigure[$\mathbf{6_{\rm II}:}$ $\mathcal{B}_{6_{\rm II}1} = -0.40 \pm 0.30$]{\includegraphics[width = 0.17\linewidth, angle=-90]{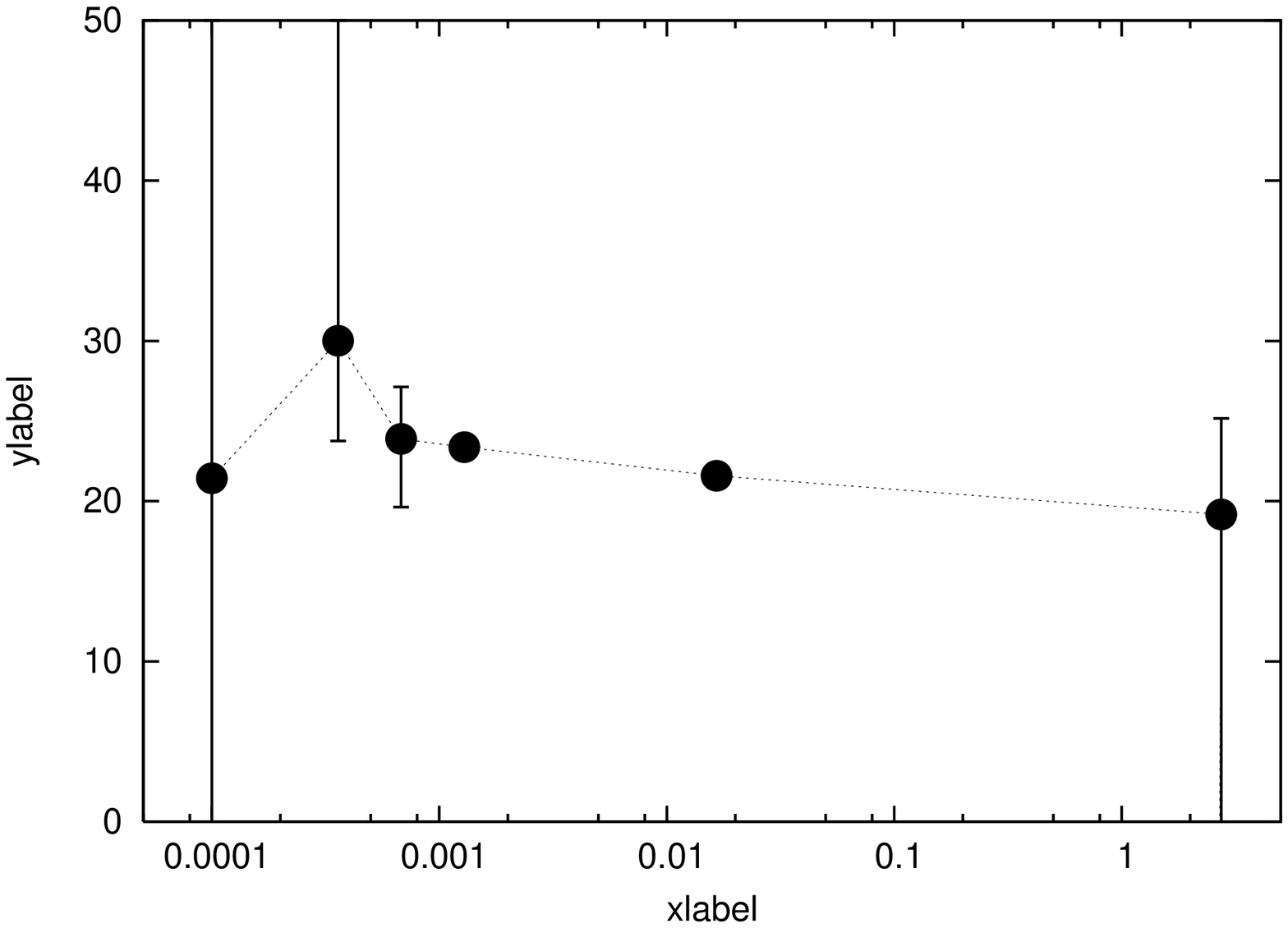}}
\subfigure[$\mathbf{6_{\rm III}:}$ $\mathcal{B}_{6_{\rm III}1} = -2.10 \pm 0.30$]{\includegraphics[width = 0.17\linewidth, angle=-90]{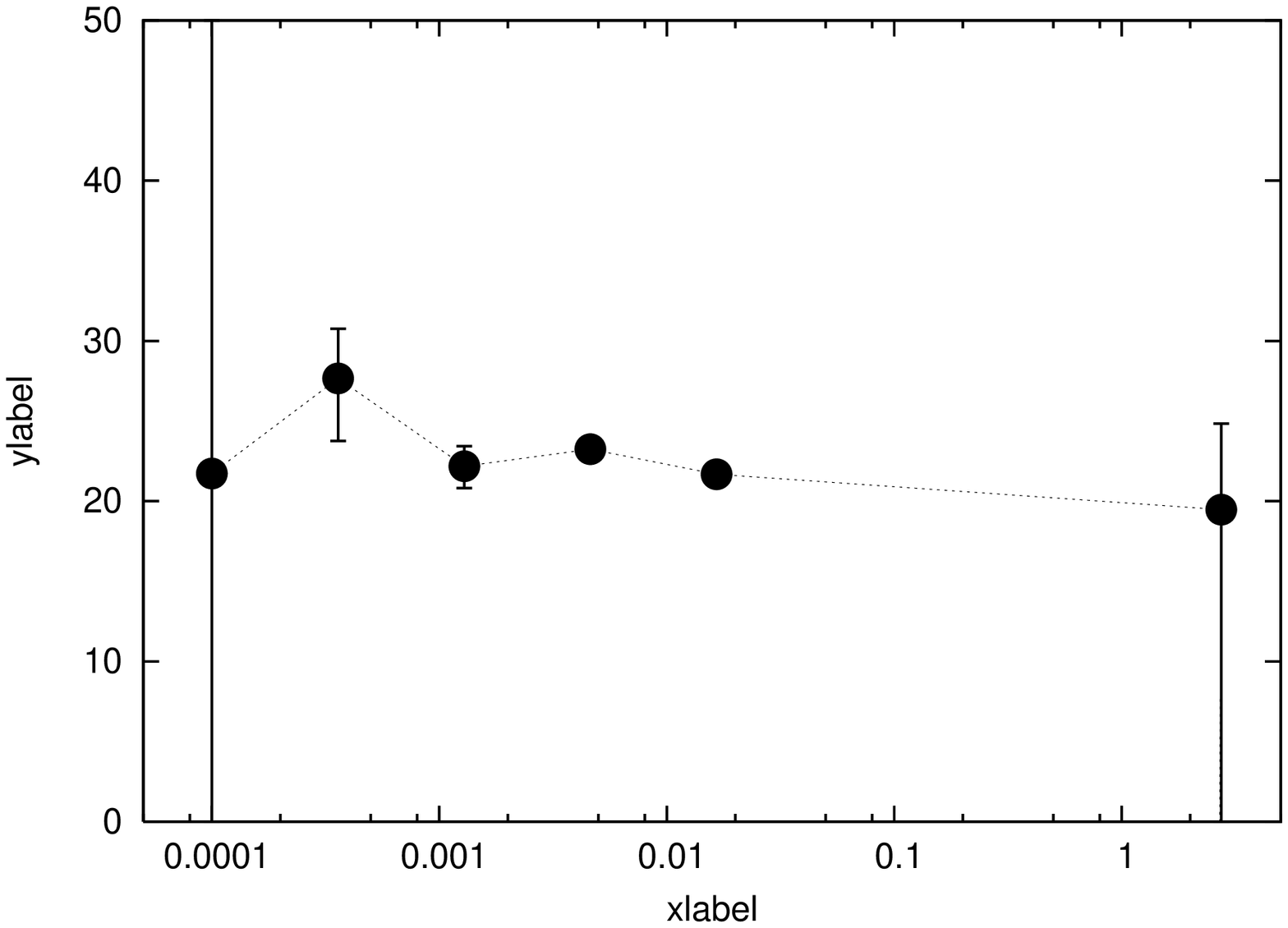}}
\subfigure[$\mathbf{6_{\rm IV}:}$ $\mathcal{B}_{6_{\rm IV}1} = -1.97 \pm 0.30$]{\includegraphics[width = 0.17\linewidth, angle=-90]{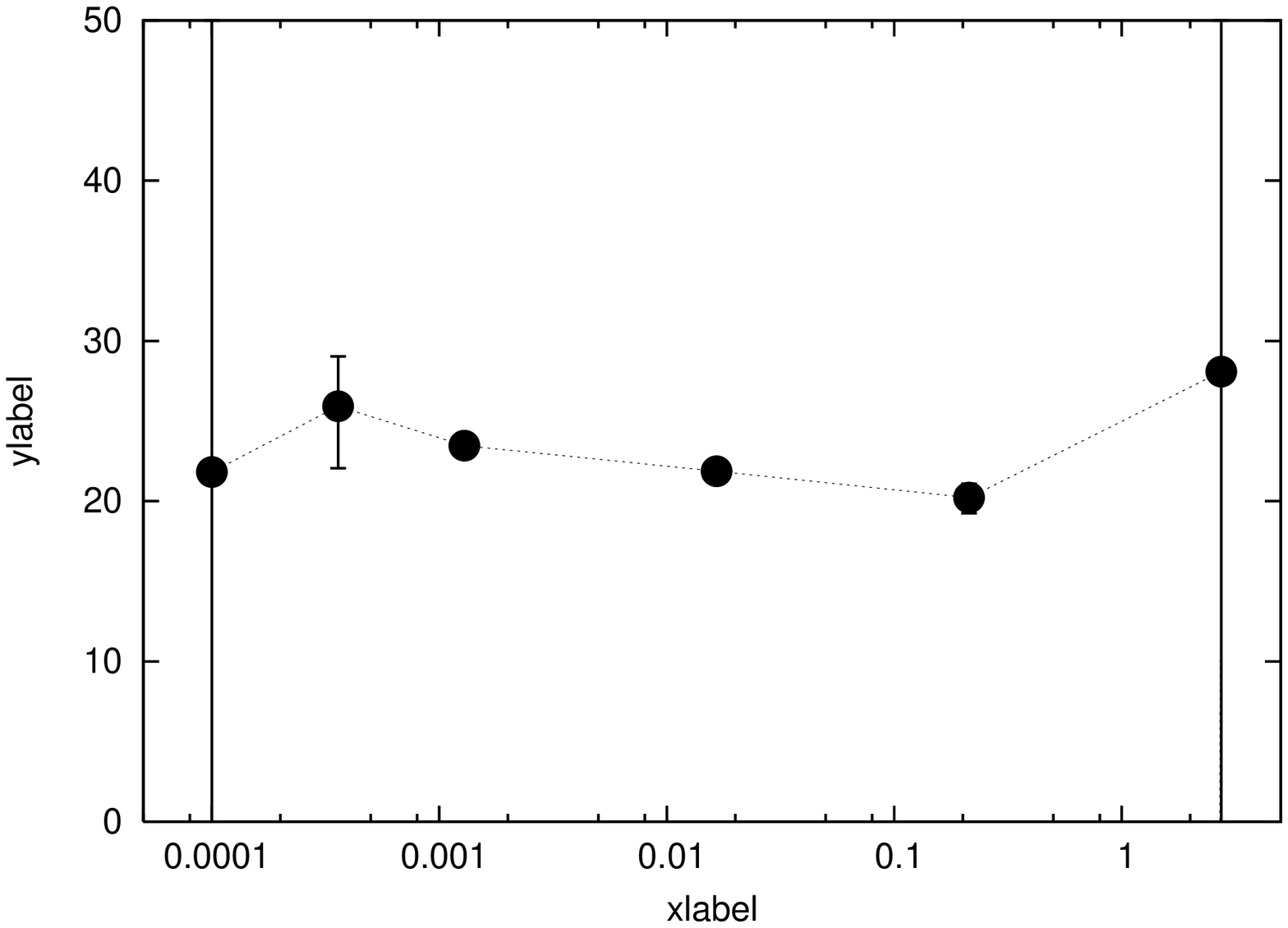}}
\label{figure:binned}
\caption{Linear interpolated reconstructions of the primordial spectrum with associated Bayes'
factors with respect model $1$. The amplitude was
allowed to vary at each of the nodes (shown with black circles). Mean amplitude values 
and
$1\sigma$ limits are shown, taken from the posteriors illustrated in
Fig.~6.}
\end{figure*}

\begin{figure*}
\psfrag{xlabel}{\tiny $\mathcal{P}(k) \times 10^{-10}$}
\psfrag{ylabel}{}
\centering
\subfigure[{\bf 1:} $\mathcal{B}_{11} = 0.00 \pm 0.30$]{\includegraphics[width = 0.17\linewidth,angle =-90]{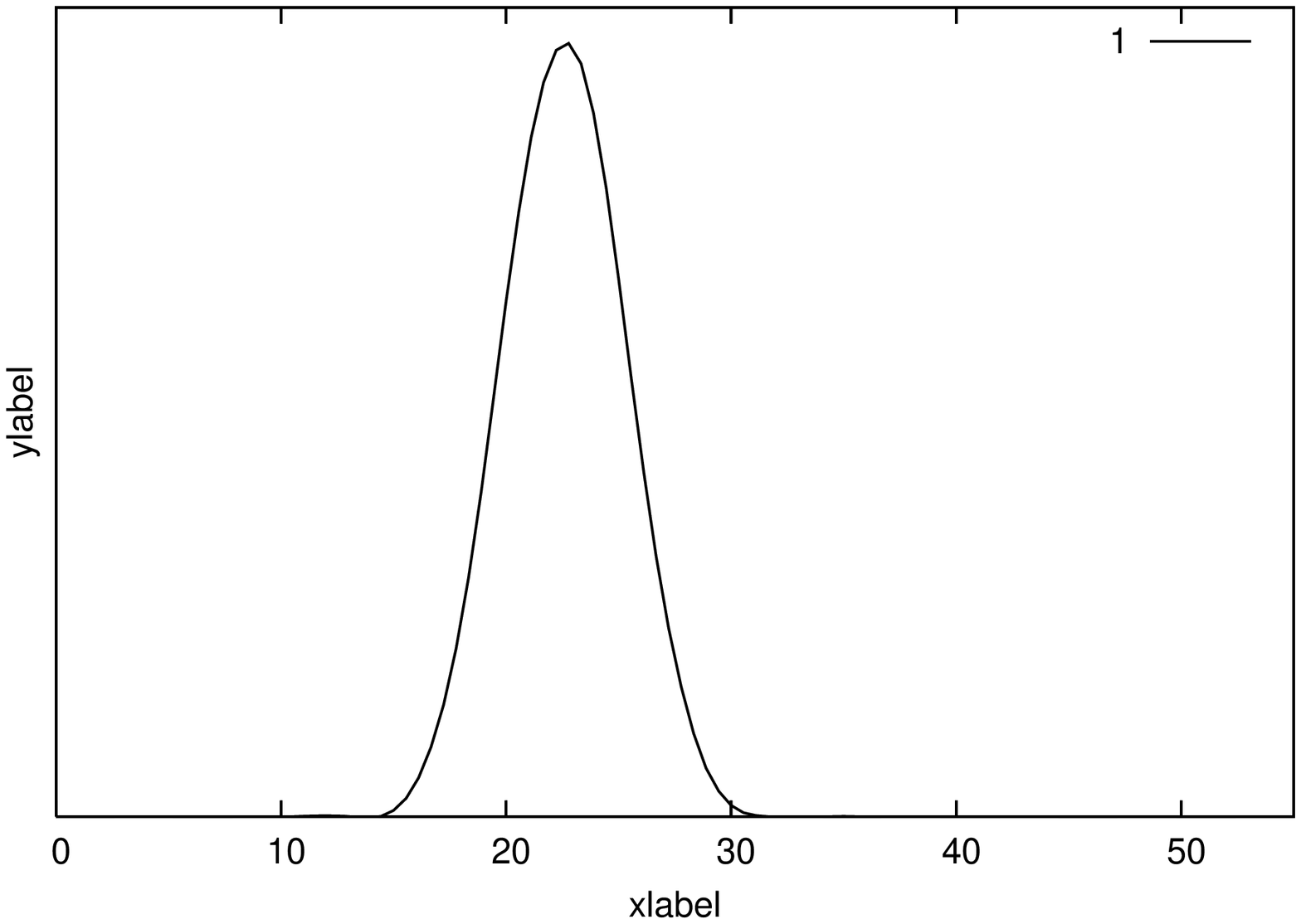}}\\
\subfigure[{\bf 2:} $\mathcal{B}_{21} = +0.66 \pm 0.30$]{\includegraphics[width = 0.17\linewidth, angle =-90]{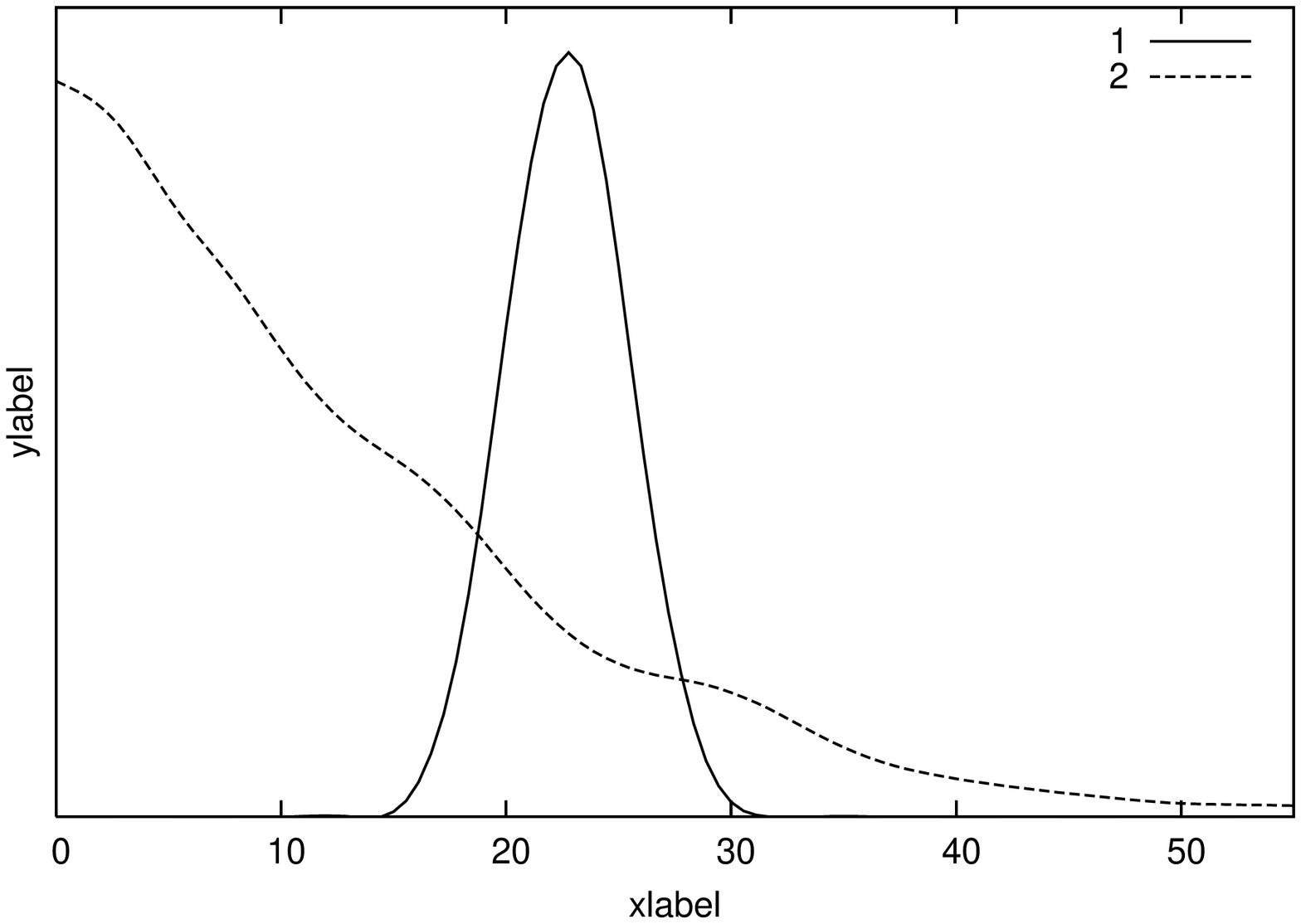}}\\
\subfigure[$\mathbf{3:}$ $\mathcal{B}_{31} = +1.08\pm 0.30$]{\includegraphics[width = 0.17\linewidth, angle =-90]{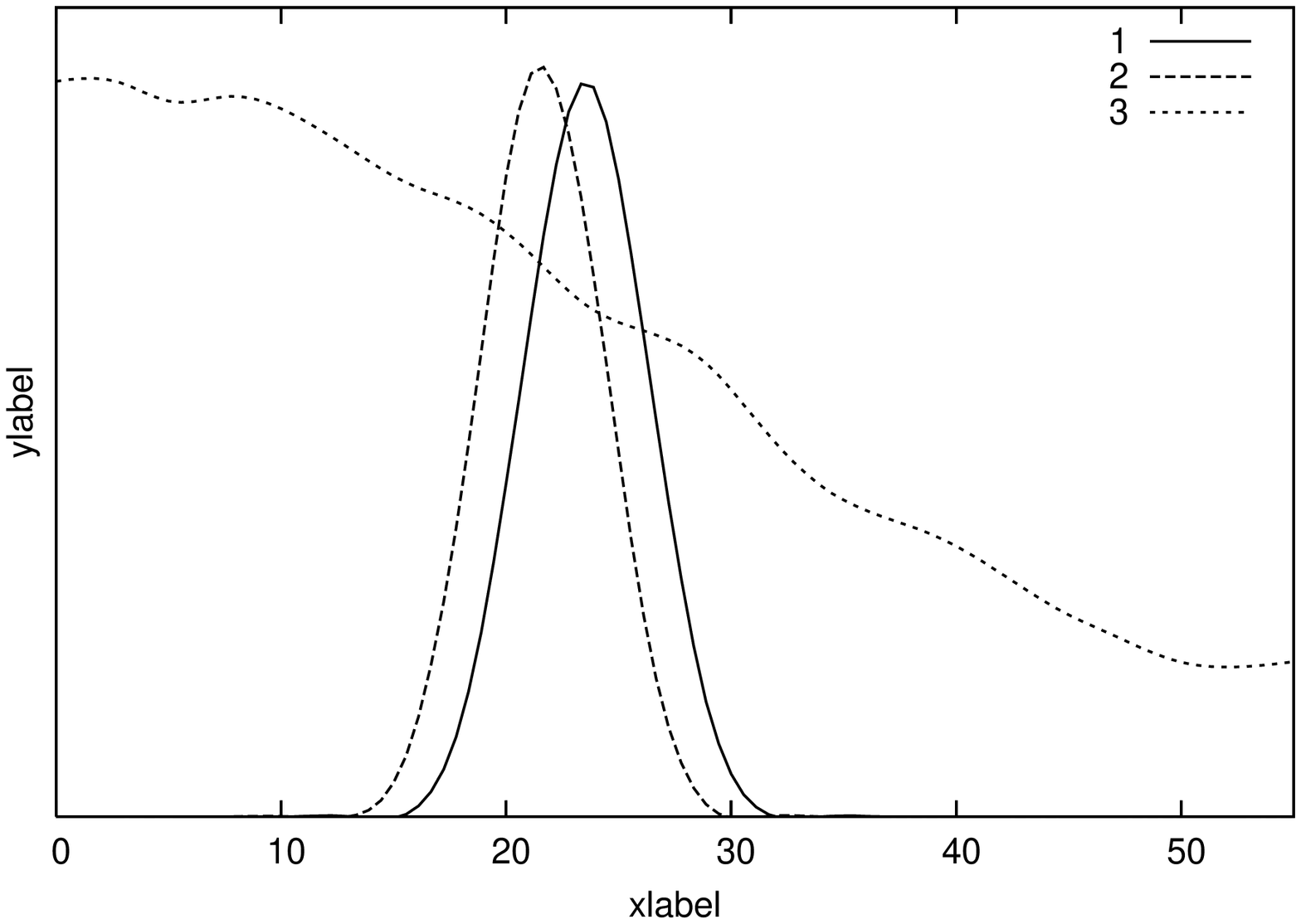}}\\
\subfigure[$\mathbf{4_{\rm I}:}$ $\mathcal{B}_{4_{\rm I}1} = -0.34 \pm 0.30$]{\includegraphics[width = 0.17\linewidth, angle=-90]{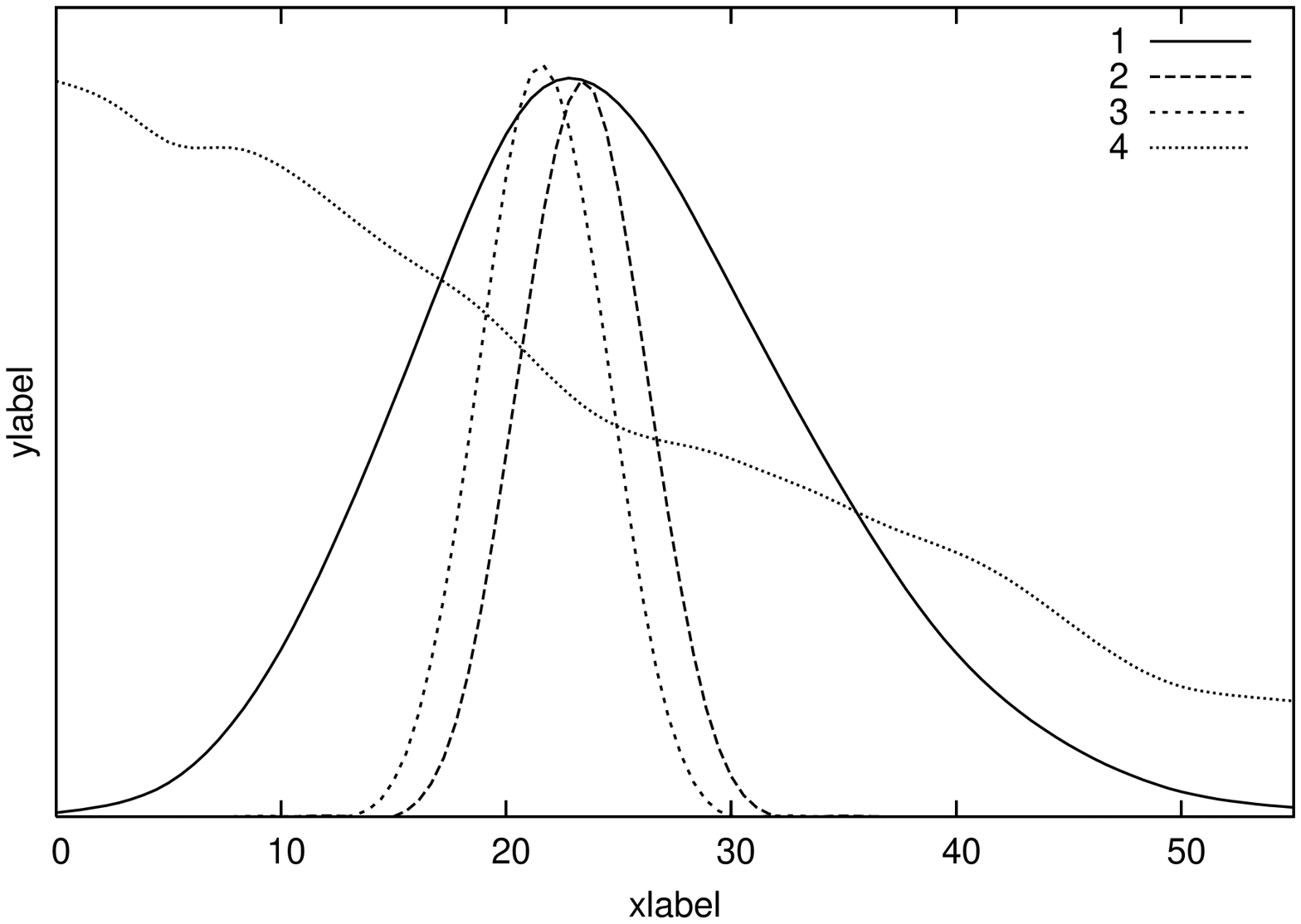}}
\subfigure[$\mathbf{4_{\rm II}:}$ $\mathcal{B}_{4_{\rm II}1} = -1.41 \pm 0.30$]{\includegraphics[width = 0.17\linewidth, angle =-90]{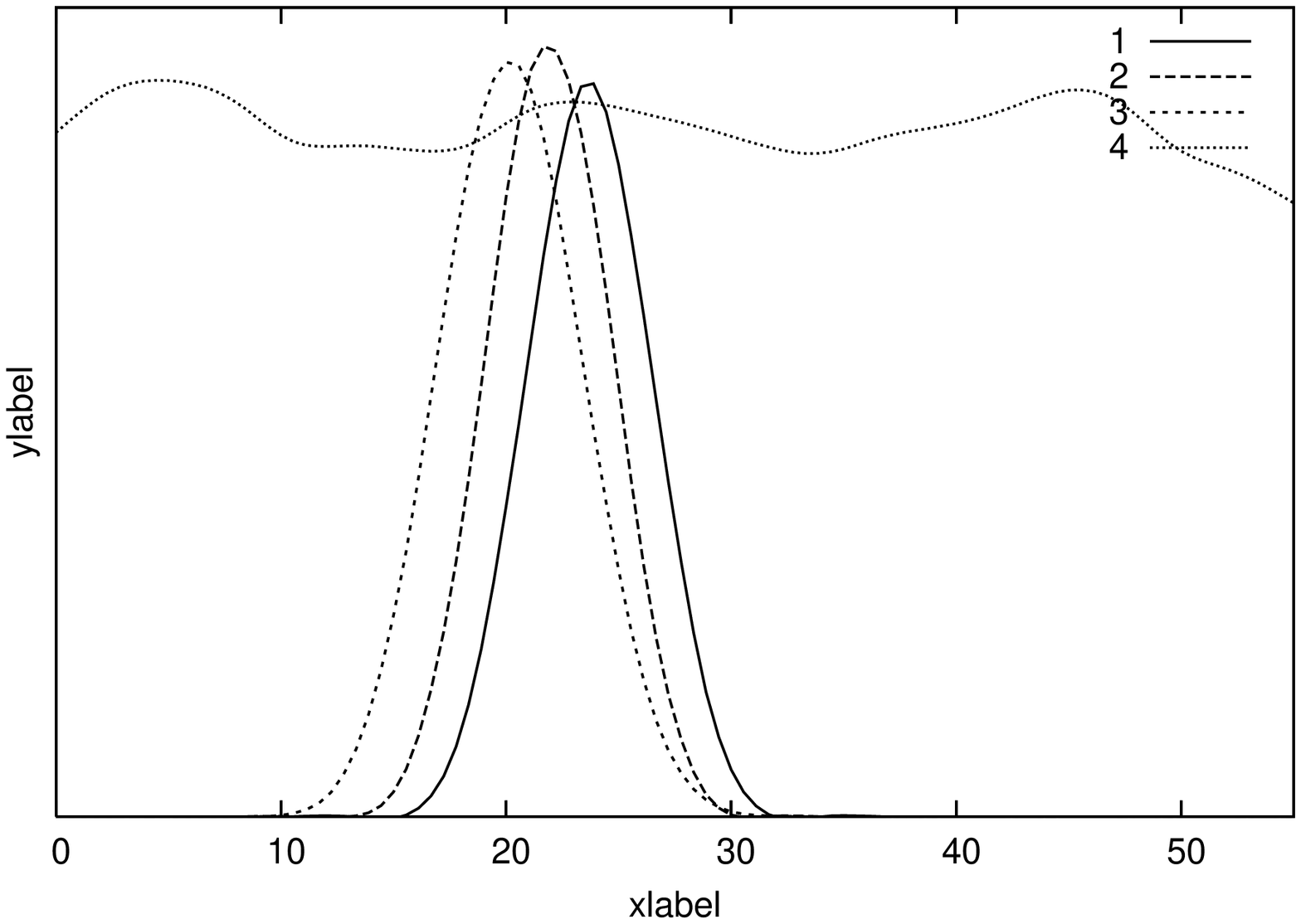}}\\

\subfigure[$\mathbf{5_{\rm I}:}$ $\mathcal{B}_{5_{\rm I}1} = -0.51 \pm 0.30$]{\includegraphics[width = 0.17\linewidth, angle=-90]{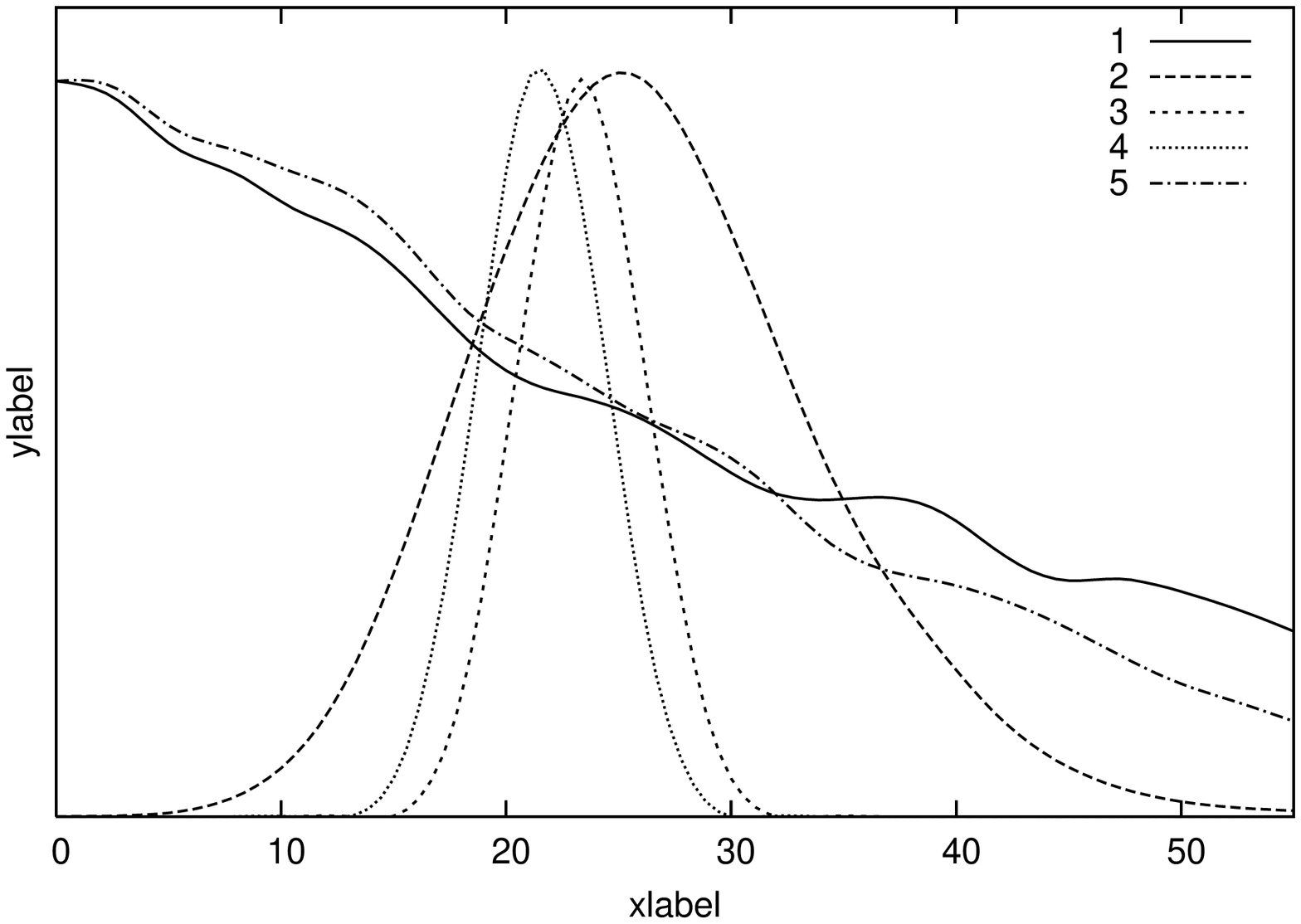}}
\subfigure[$\mathbf{5_{\rm II}:}$ $\mathcal{B}_{5_{\rm II}1} = -2.41 \pm 0.30$]{\includegraphics[width = 0.17\linewidth,angle=-90]{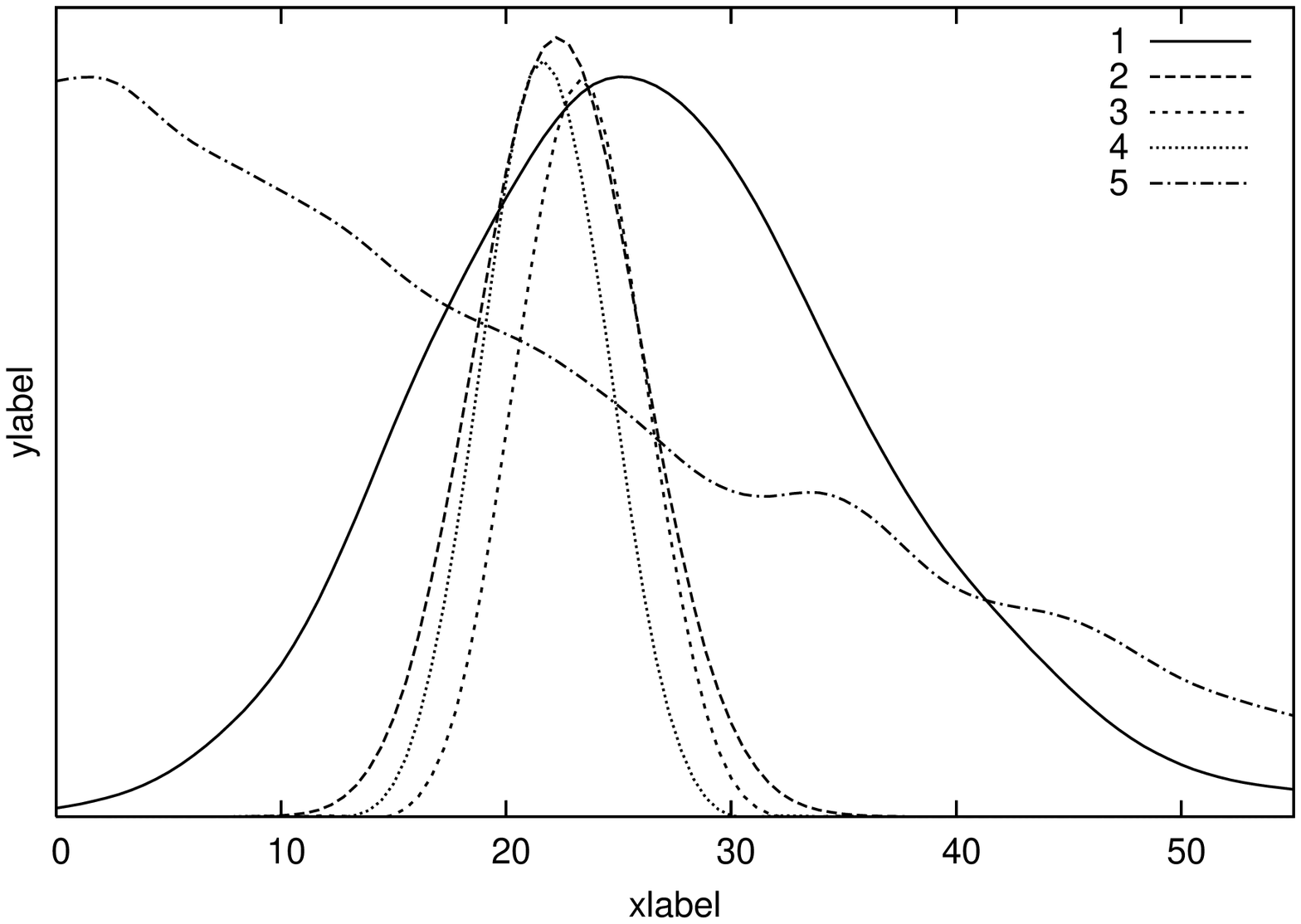}}
\subfigure[$\mathbf{5_{\rm III}:}$ $\mathcal{B}_{5_{\rm III}1} = -2.05 \pm 0.30$]{\includegraphics[width = 0.17\linewidth, angle =-90]{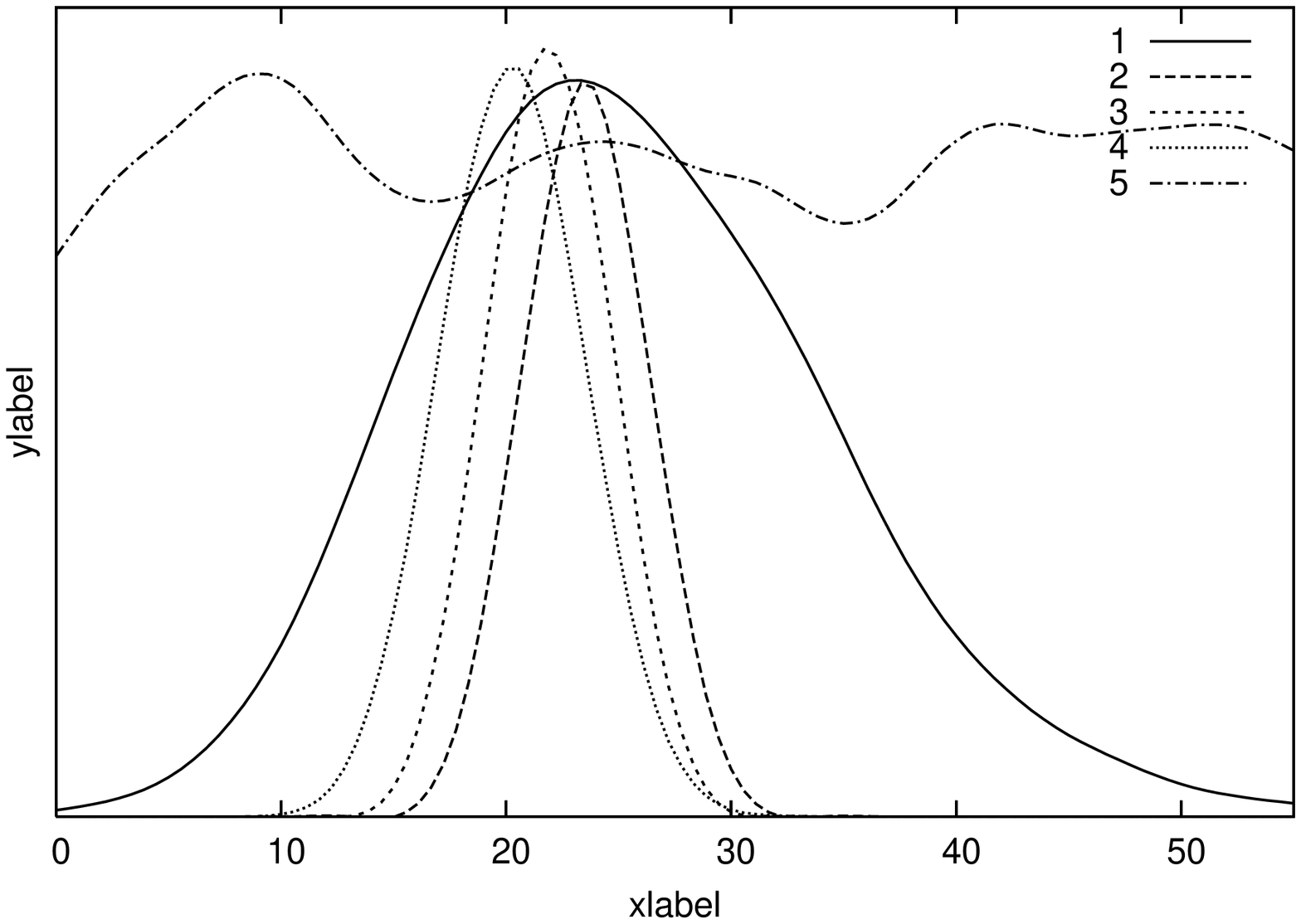}}\\

\subfigure[$\mathbf{6_{\rm I}:}$ $\mathcal{B}_{6_{\rm I} 1 } = -0.21 \pm 0.30$]{\includegraphics[width = 0.17\linewidth, angle=-90]{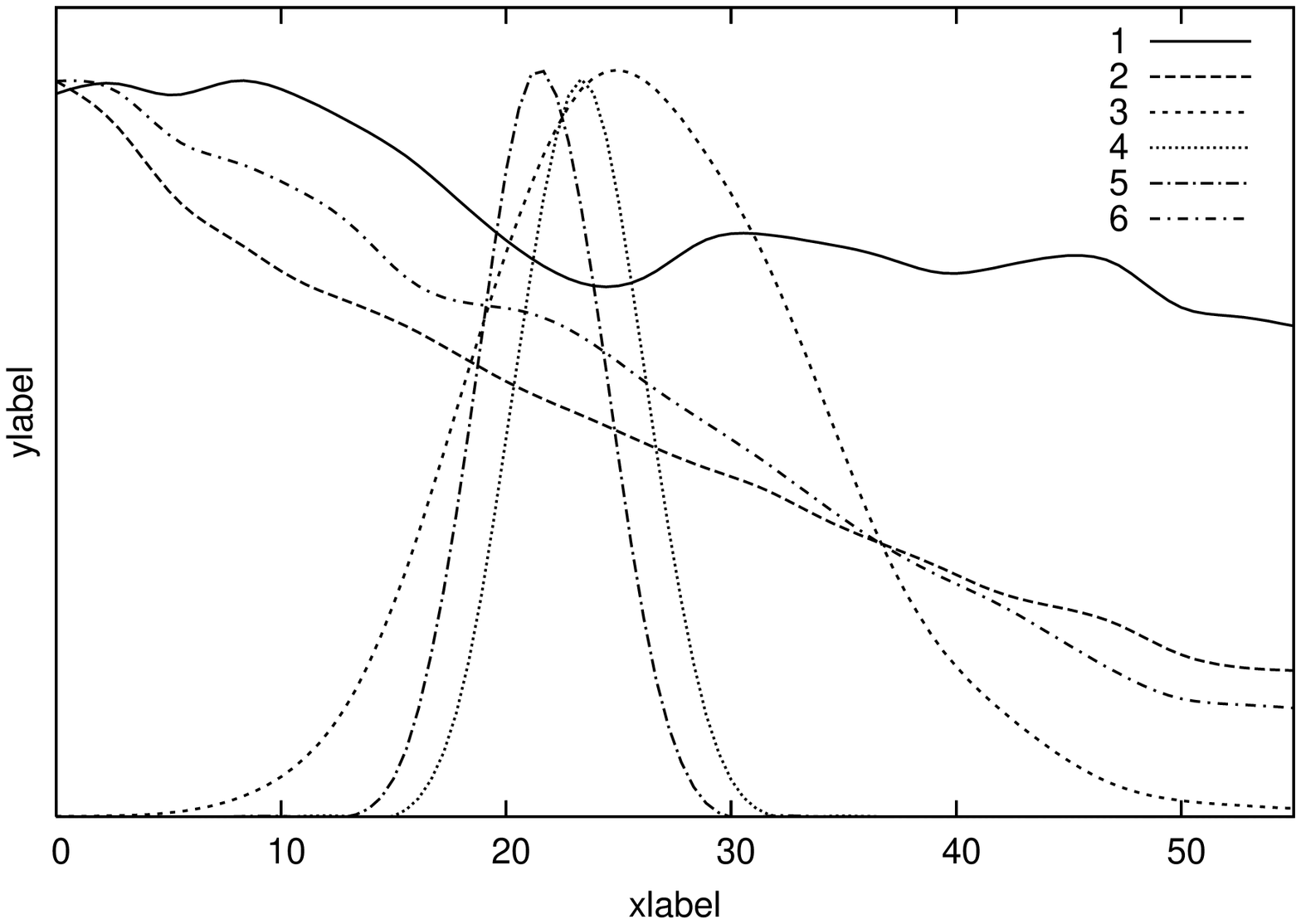}}
\subfigure[$\mathbf{6_{\rm II}:}$ $\mathcal{B}_{6_{\rm II}1} = -0.40 \pm 0.30$]{\includegraphics[width = 0.17\linewidth, angle=-90]{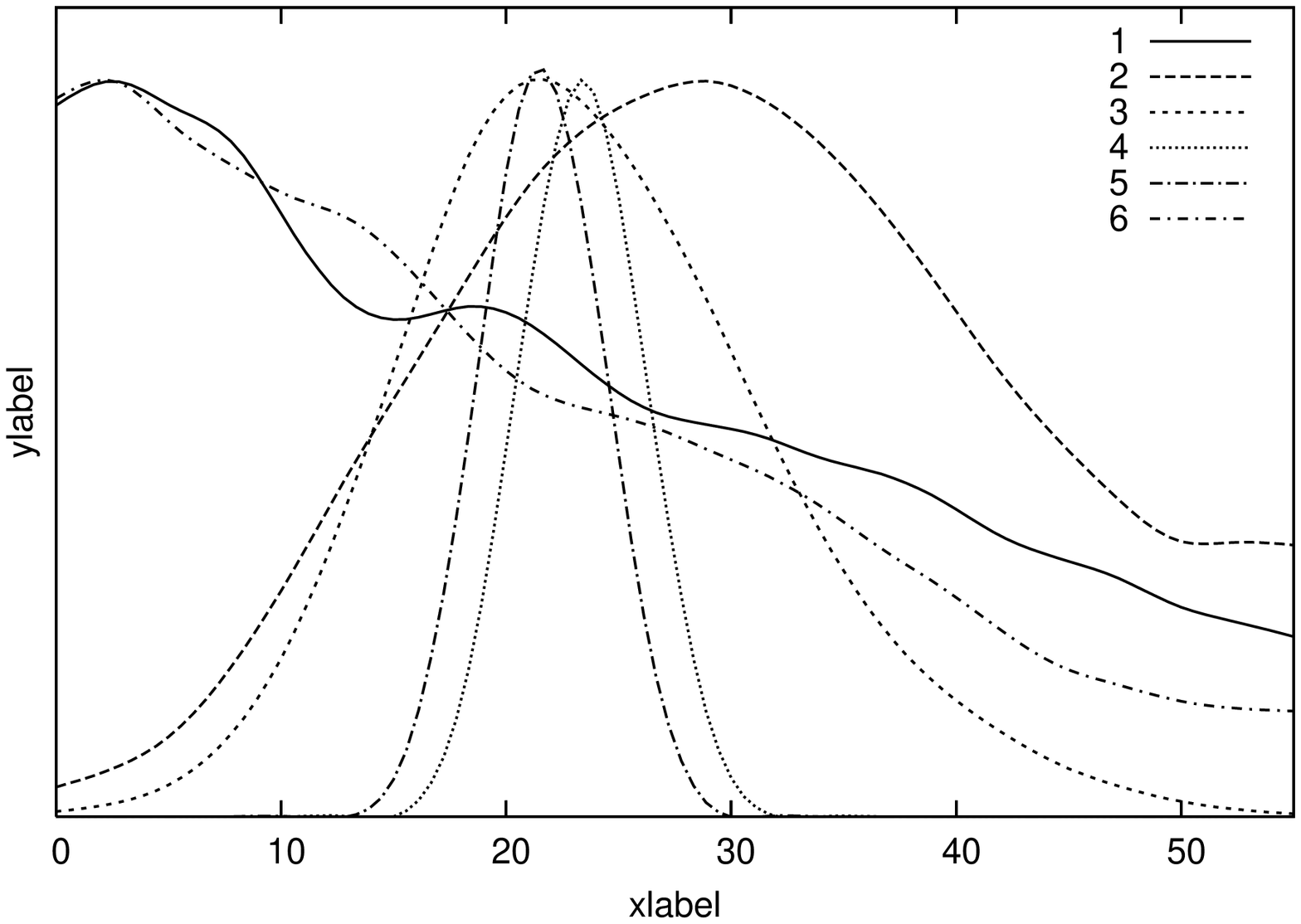}}
\subfigure[$\mathbf{6_{\rm III}:}$ $\mathcal{B}_{6_{\rm III}1} = -2.10 \pm 0.30$]{\includegraphics[width = 0.17\linewidth, angle=-90]{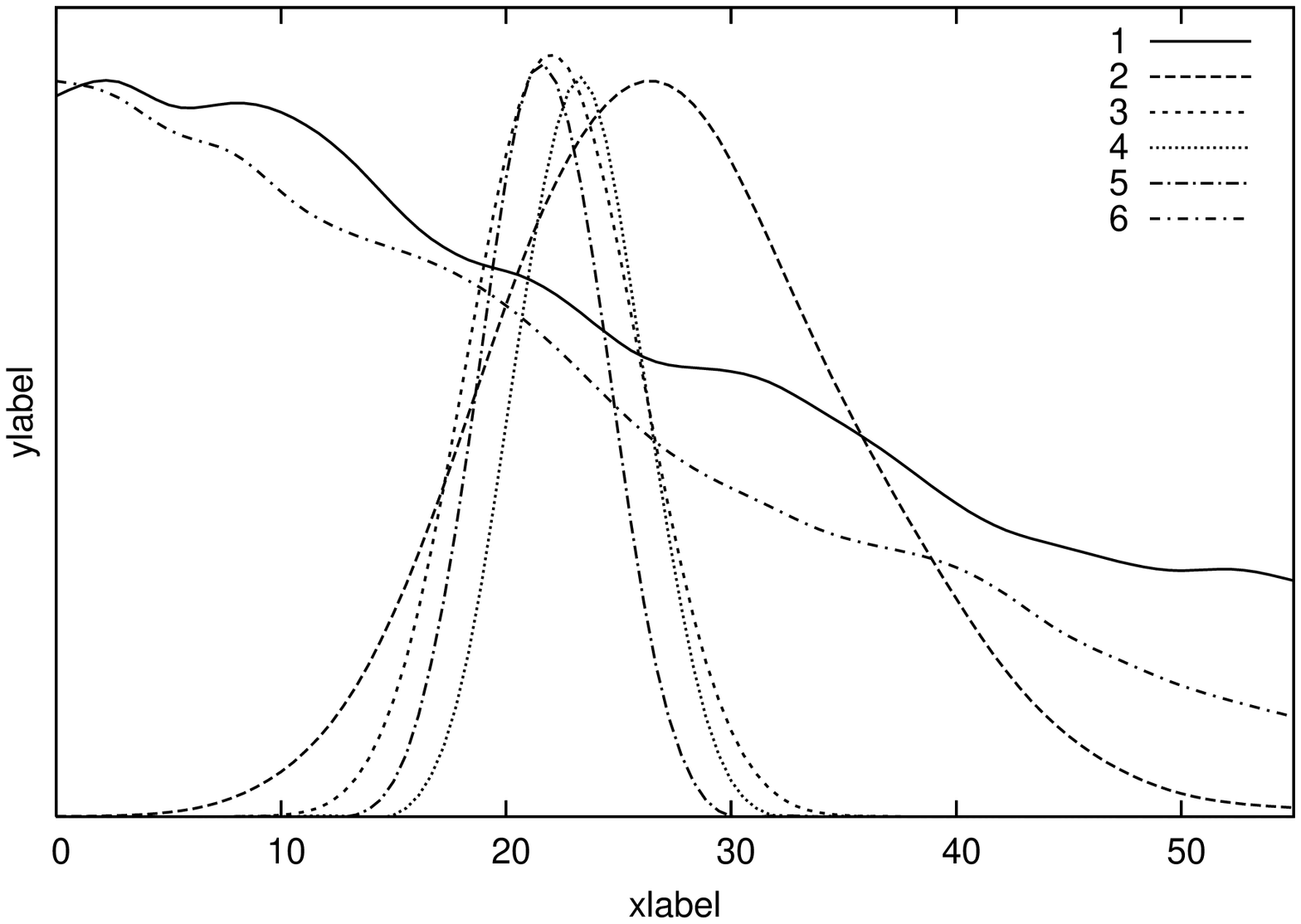}}
\subfigure[$\mathbf{6_{\rm IV}:}$ $\mathcal{B}_{6_{\rm IV}1} = -1.97 \pm 0.30$]{\includegraphics[width = 0.17\linewidth, angle=-90]{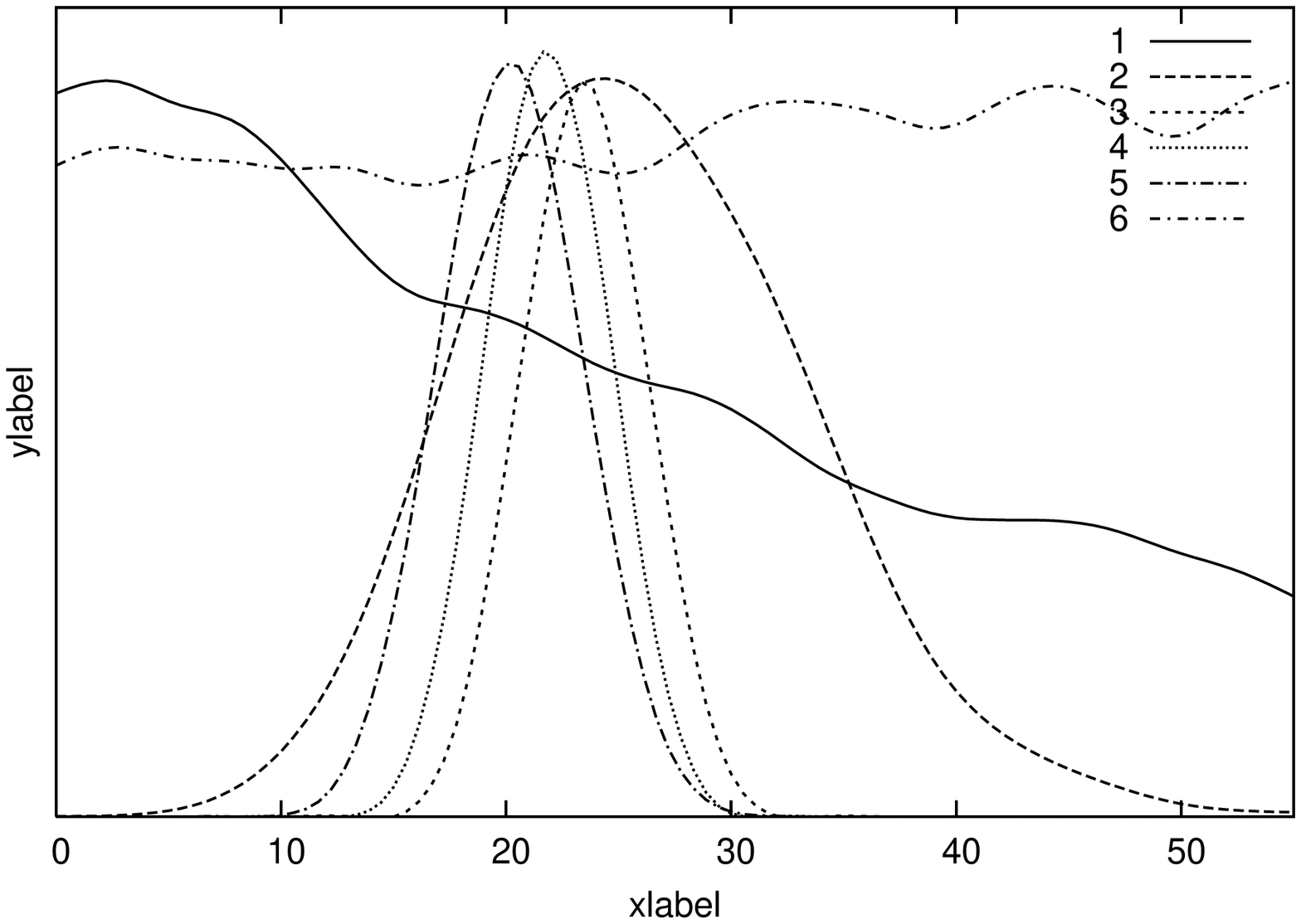}}
\label{figure:binned_posteriors}
\caption{Marginalised 1-dimensional posterior distributions of the amplitude at each
$k$-space node  
used in each reconstruction.}
\end{figure*}

\subsection{Model Comparison I: the Bayesian evidence}
The marginalised 1-dimensional posterior distributions for the amplitude at each node
 and for each
reconstruction are shown in Fig.~6
[Fig.~5 illustrates the corresponding form of the reconstructed spectra from
the mean posterior estimates (with $1\sigma$ error bars on the amplitudes)].  Comparing figure
(b) in both Figs. 5 and 6 we see that there is only
an upper bound on the amplitude at $k_{\rm max}$, with no lower bound. This is a consequence of our
choice of a large $k_{\rm max}$, well above any current experimental constraint and simply
allows the power to gradually fall to zero.  
The difference in evidence is minimal between the
base and two node model with $\mathcal{B}_{12}=0.66$ being too small, on the Jeffreys' scale
to draw any decisive conclusions, though within the error the evidence marginally prefers model
$2$. The third model adds a node at $k\sim 0.0166$ Mpc$^{-1}$ emulating a degree
of spectral running by allowing a slight variation in the interpolated slopes between the
three nodes. Though no meaningful constraint is possible at the upper $k$ scale, this model is
preferred over model $2$ with $\mathcal{B}_{23} \sim 0.4$ and significantly over the base 
model by
$\mathcal{B}_{02} \sim 1.1$ units.
The fourth stage reconstruction requires us to test two combinations of
node positions, the first, $4_{\rm I}$, splits the lowest $k$ \emph{bin} at $k\sim0.00129$
Mpc$^{-1}$ while the second, $4_{\rm II}$ divides the upper $k$ bin. 
$\mathcal{B}_{3 4_{\rm I}}$ and 
$\mathcal{B}_{3 4_{\rm II}}$ both significantly disfavour the addition of a fourth node. 
This result points to some deviation from scale invariance at around the position 
$k \sim 0.01$, the rough location of the additional node in model 3. Further parameterisation
both above ($4_{\rm II}$) and below ($4_{\rm I}$) this scale is disfavoured, 
lending credence to the
general conclusions of \citet{Verde07} who found a similar `turn-over' scale. According to
the evidence the optimal reconstruction contains, perhaps surprisingly only three parameters.


It is
interesting to note that the parameterisation in $4_{\rm I}$ is significantly preferred over
$4_{\rm II}$, i.e. an additional node seems to be preferred on large scales over small.
Although
technically redundant we can continue to a fifth and sixth stage to see if this effect 
continues. Assuming then that the fourth stage evidence has now indicated
a preference for large scale (small $k$) structure over small  we continue by
sub-dividing the largest $k$ bin of $4_{\rm I}$ again at $k\sim0.00036$, which we denote as 
$5_{\rm I}$. The two other possible splittings being $5_{\rm II}$ at $k\sim0.00462$ and
$5_{\rm III}$ at $k\sim0.21$. To within estimated error $\mathcal{B}_{4_{\rm I} 5_{\rm I}} \sim 0$ and
again both $5_{\rm II}$ and $5_{\rm III}$ are significantly disfavoured. This result is 
repeated at the sixth stage.    

So, curiously, although the evidence peaks at model $3$ 
there is a
substantial preference in all subsequent 
reconstructions for additional amplitude nodes to be placed at large scales 
(i.e. models $4_{\rm I}$, $5_{\rm I}$ and $6_{\rm I}$). Furthermore the evidence is observed to   
\emph{plateau} in value with $\mathcal{B}_{4_{\rm I} 5_{\rm I}}$ and $\mathcal{B}_{5_{\rm I} 
6_{\rm I}}$ being
roughly zero. 
The first result could suggest that although the data cannot yet cope with the
extra complexity, large scale structure is useful in a model. However when combined with the
second result this points to the additional parameters not \emph{over complicating} the 
model but instead being \emph{ignored} and left unconstrained by the data.     
 The evidence is quite deliberately adept at
ignoring such extra complexity; the extra undetermined parameter direction simply does not
affect the average posterior over the prior. This effect is demonstrated here by comparing
Figures~5 (d) and (f) where the act of placing an additional node at
$\sim0.00129$ Mpc$^{-1}$ removes all constraint on the amplitude at node $k_{\rm min}$ and
thus \emph{de-facto} removes a parameter from the analysis. To account correctly for this
effect, the analyst requires a further level of model discrimination, that can interpret
quantitatively the constraining power of a given model and data combination. For this we must
fully define what we are penalising in extra model
complexity, and for this we turn to the Bayesian \emph{complexity}. 

\subsection{Model Comparison II: the Bayesian complexity}
The advantage of Bayesian model selection is that it penalises model parameters 
that cannot be justified by the data. However the number of free parameters is 
only the most naive measure of the complexity of a model. A more thorough comparison 
can be gleaned from what is termed the Bayesian or \emph{effective} complexity of a model.
This definition was first given by \citet{Spiegelhalter} and was subsequently introduced into
cosmology by \citet{Kunz}. The starting point is a quantifiable definition of how a set
of data can improve the prior knowledge of a model. In other words a measure of the
relative difference between the
posterior and prior distributions, sometimes termed the \emph{information gain}. 
The Kullback-Leibler (KL) divergence $D_{\rm KL}$ measures just this, via the relative
entropy between two probability distributions, $P$ and $\pi$: 
\begin{equation}
D_{\rm KL} (P, \pi) \equiv \int P(\mathbf{\Theta}|\mathbf{D}) \ln 
\frac{P(\mathbf{\Theta}|\mathbf{D})}{\pi(\mathbf{\Theta})} {\rm d} \mathbf{\Theta}.
\label{equation:divergence}
\end{equation}
From this definition the Bayesian complexity can then be defined as the difference in 
$D_{\rm KL}$ between some real experiment and the ideal situation where the information
gain is maximised.  
To see how this works, let us take the ideal example of a uniform prior distribution $\pi$ and an
excellent set of data $\mathbf{D}$ such that on completion of a Bayesian analysis the prior
distribution collapses into a $\delta-$function posterior distribution about some parameter
vector $\mathbf{\Theta '}$. This we take as our \emph{ideal} scenario in which the divergence
between posterior and prior is maximised and is given approximately by $D_{\rm KL}' =
\ln P(\mathbf{\Theta '})/ \pi(\mathbf{\Theta '})$. In a realistic experiment of course 
the posterior
$P(\mathbf{\Theta})$ will resemble some (approximately) multidimensional  Gaussian
distribution with some mean $\widehat{\mathbf{\Theta}}$ parameter vector and an associated
variance so that the divergence would be given simply by Eqn. \ref{equation:divergence}. 
The Bayesian complexity $C_{\rm B}$ can thus be defined as the difference between the ideal 
point estimate $D_{\rm KL}'$ and the actual divergence:
\begin{equation}
C_{\rm B} \equiv -2\left( D_{\rm KL} (P, \pi) - \widehat{D_{\rm KL}} \right). 
\label{equation:complexity}
\end{equation}
This leaves us free to choose an
appropriate point estimate that maximises information gain -which for most well constrained 
cosmological
problems can be taken to be the \emph{mean} of the full posterior distribution. 
Using Eqn.~\ref{equation:divergence} and Bayes' theorem one can rewrite 
Eqn.~\ref{equation:complexity} as:
\begin{equation}
C_{\rm B} \equiv -2 \int P(\mathbf{\Theta} | \mathbf{D}) \ln \mathcal{L}(\mathbf{\Theta}) 
+ 2 \ln \mathcal{L} (\widehat{\mathbf{\Theta}}) {\rm d} \mathbf{\Theta}.
\end{equation}
By defining an effective $\chi^2$ through $\mathcal{L}(\mathbf{\Theta}) \propto e^{-
\chi^2/2}$, such that all constant factors within the likelihoods drop out, we can
define the Bayesian complexity as:
\begin{equation}
C_{\rm B} = \overline{\chi^2 (\mathbf{\Theta})} - \chi^2 (\overline{\mathbf{\Theta}}),
\end{equation}
where the first term denotes the mean $\chi^2$ across a set of posterior samples 
while the second term is the $\chi^2$ at the mean parameter
values. 

Based on this definition the Bayesian complexity succinctly compares the  \emph{constraining
power} of the data with the \emph{predictivity} of the model. Thus a model with highly
restrictive priors, and unconstrained posteriors will have a \emph{low} Bayesian
complexity, as the predictiveness of the model was already very high initially.
Conversely, wide priors with highly constrained posteriors will result in a \emph{high}
complexity (which can tend to a maximal value equal to the \emph{actual} number of model 
parameters, $C_{\rm 0}$) 
as the data constrained the model substantially over 
the uninformative priors.  

\begin{table}
\begin{center}
\caption{The reconstruction Bayesian complexity $C_{\rm B}$ and actual number of model parameters
$C_{\rm 0}$.}
\begin{tabular}{|c||c||c|}
    \hline
 \textbf{Model} &  $C_{\rm 0}$ 	& $C_{\rm B}$ \\
    \hline
 $1$ 		&  $7$ 		& $5.35 \pm 0.10$\\
 $2$ 		&  $8$ 		& $6.35 \pm 0.10$\\
 $3$ 		&  $9$ 		& $7.03 \pm 0.10$\\
 $4_{\rm I}$ 		&  $10$ 	& $7.82 \pm 0.10$\\    
 $4_{\rm II}$ 	&  $10$ 	& $7.18 \pm 0.10$\\  
 $5_{\rm I}$ 		&  $11$ 	& $8.04 \pm 0.10$\\    
 $5_{\rm II}$ 	&  $11$ 	& $8.60 \pm 0.10$\\ 
 $5_{\rm III}$ 	&  $11$ 	& $8.37 \pm 0.10$\\     
 $6_{\rm I}$ 		&  $12$ 	& $8.04 \pm 0.10$\\    
 $6_{\rm II}$ 	&  $12$ 	& $8.60 \pm 0.10$\\ 
 $6_{\rm III}$ 	&  $12$ 	& $8.37 \pm 0.10$\\
 $6_{\rm IV}$ 	&  $12$ 	& $8.37 \pm 0.10$ \\ 
    \hline
\end{tabular}
\label{table:complexity}
\end{center}
\end{table}

\begin{figure}
\begin{center}
\psfrag{xlabel}{$C_{\rm 0}$}
\psfrag{ylabel}{$C_{\rm B}$}
\includegraphics[width=0.6\linewidth, angle = -90]{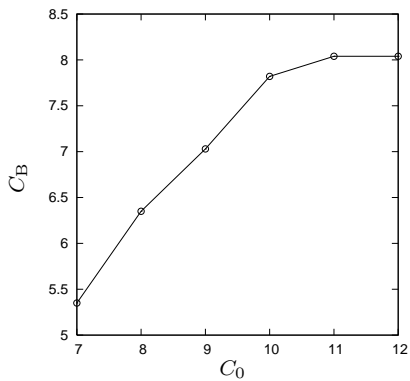}
\caption{Bayesian complexity $C_{\rm B}$ versus actual number of model parameters $C_{\rm 0}$ 
for models: 1, 2, 3, $4_{\rm I}$, $5_{\rm I}$ and $6_{\rm I}$. Note how 
$C_{\rm B}$ increases almost linearly with $C_{\rm 0}$ until model $4_{\rm I}$ ($C_{\rm 0} = 10$)
when $C_{\rm B}$ begins to plateau in value as successively less well constrained parameters are added.}
\label{figure:complexity}
\end{center}
\end{figure}

It should be emphasised that estimates of the Bayesian complexity cannot be used in isolation
for model selection, blindly choosing the model with the smallest complexity would simply
\emph{under-fit} the data. Instead it provides a useful discriminator in cases where the
evidence difference between models is so small (say $<$ $1$ log unit on the Jeffreys' scale)
that little inference can be drawn with the evidence alone. Besides the most obvious
scenario, where both models are essentially equally informative, the case, as we had in the
last section can be envisaged where additional parameters are simply left unconstrained by
the data, such that in the evidence integral this direction is simply \emph{averaged over}.
Here the complexity can quantify whether or not the additional parameters have actually 
been constrained and thus extracted any further information from the data.  

Table~\ref{table:complexity}
lists the recovered complexity for each of our reconstructions tested. It should be noted
that we have chosen quite a generic background cosmology accounting for both the possibility
of spatial curvature, via the $\Omega_k$ parameter and the marginalisation over a possible SZ
contribution at high $\ell$ as was done in \citet{Komatsu}. Inclusion of recent LRG data with
their associated tight constraints on $\Omega_k$ will minimise any effect on $C_{\rm B}$,
however $A_{SZ}$ remains essentially unconstrained by current data. Thus it is not surprising
to see our base, scale invariant model $1$ having an effective complexity significantly less
than $C_{\rm 0}$. This need not concern us here however, as we are primarily interested in
the relative difference of $C_{\rm B}$ as we increase the reconstruction complexity. 

Since
the evidence is maximised for model $3$, this should be our preferred parameterisation. Of
course the Bayes' factor $\mathcal{B}_{32}$ between models $3$ and $2$ is only $\sim 0.4$, or
on the Jeffreys scale of little significance, and since the Bayesian complexity for model $2$ is
significantly smaller (by $\sim 0.7$) than $3$ should we then argue that model $2$ should in
fact be preferred?
Looking at the marginalised posteriors of $3$, the fact that it is preferred is not at all surprising,
as the addition of the node at $k\sim 0.01656$ Mpc$^{-1}$ 
leaves no amplitude constraint at $k_{\rm max}$.
In effect the evidence is maximised for model $3$ as it is a
\emph{de-facto} two parameter model. However crucially it provides the required tilt
over a $k$ range that is well constrained by data \emph{and} allows a deviation in this 
tilt above $k \sim 0.01$. Further modelling of the upper tilt, via say an extra node 
as we performed in model ${4_{\rm II}}$ was strongly disfavoured $\mathcal{B}_{3 4_{\rm
II}} \sim 2.5$. So the inclusion of complexity in the analysis has not altered our
general conclusions, as the evidence difference between models 2 and 3 is minimal,
it simply serves to highlight the lack of significance placed by the data in anything other
than a tilted spectrum at present.

The complexity can further explain the degneracy in evidence values for those models 
where we introduced additional large scale structure (e.g. $4_{\rm I}$, $5_{\rm I}$ and $6_{\rm I}$).
Fig.~\ref{figure:complexity} plots $C_{\rm B}$ against $C_{0}$ for these models (and
for comparison the first three models). As we increase the number of parameters in
going from model 1 to 3 the Bayesian complexity is seen to rise roughly linearly, from
which we infer that the data can usefully constrain all of the model parameters and
thus can warrant the additional parameterisation. 
This trend continues to model $4_{\rm I}$, but thereafter $C_{\rm B}$ tends rapidly to 
a constant value of $\sim 8$, suggesting that the inclusion of extra parmeters in 
models $5_{\rm I}$ and $6_{\rm I}$ is superfluous. 
Thus despite the indifference
shown by the evidence the Bayesian complexity has successfully, and correctly, 
relegated these models.

\section{Conclusions}
\label{section:conclusions}
In this paper we have attempted to fit an optimal degree of structure to the primordial power
spectrum using Bayesian model selection tools as our discriminant criteria. We find that a scale
invariant spectrum is significantly ruled out, 
the data instead favouring a tilted spectrum, with perhaps some
slight scale dependence of $n_{\rm s}$ located close to $k \sim 0.01$ Mpc$^{-1}$. We fail to find 
any support in the data for further
features beyond this simple scenario, the optimal reconstruction fitting between just two and
three parameters. 
Previous authors (including ourselves) 
have
regularly used many more degrees of freedom, finding a number of
`interesting' features in the process. In this analysis, by accounting for Occams' razor 
we have found
no statistically significant structure, much beyond a simple tilt, and there is, we feel, 
limited 
point in attempting more complex models at present, as the data simply cannot support them.

\section*{Acknowledgements}

This work was carried out largely on the Cambridge High Perfomance Computing Cluster, 
{\sc Darwin} and we thank Stuart Rankin for assistance. We thank
Roberto Trotta for extremely helpful discussions. FF is supported by
fellowships from the Cambridge Commonwealth Trust and the Pakistan
Higher Education Commission. MB is supported by STFC. 

\bibliographystyle{mn2e}
\bibliography{paper}

\appendix

\label{lastpage}

\end{document}